\documentclass[trackchanges,twocolumn]{aastex701}

\usepackage{graphicx}
\usepackage{amsmath}
\usepackage{subcaption}
\usepackage{caption}

\submitjournal{ApJ}

\begin{document}

\title{Stationary Resistive MHD Solutions Near an Isolated Three-Dimensional Magnetic Null}

\author[0009-0001-9712-4653,gname='Kuljeet Singh Baldev Singh',sname=Saddal]{Kuljeet Singh Saddal}
\affiliation{Astronomical Institute, Czech Academy of Science, Fri\v{c}ova 298, 251 65 Ond\v{r}ejov, Czech Republic}
\affiliation{Astronomical Institute, Faculty of Mathematics and Physics, Charles University, V Hole\v{s}ovick\'{a}ch 2, 182 00 Prague, Czech Republic}
\email[show]{kuljeet.saddal@gmail.com}  

\author[0000-0001-5165-6331,gname='Dieter H.',sname=Nickeler]{Dieter H. Nickeler}
\affiliation{Astronomical Institute, Czech Academy of Science, Fri\v{c}ova 298, 251 65 Ond\v{r}ejov, Czech Republic}
\email{dieter.nickeler@asu.cas.cz}

\author[0000-0003-0346-7338,gname=Rodrigo,sname='Meneses Pacheco']{Rodrigo Meneses Pacheco}
\affiliation{Escuela de Ingeniería Civil, Universidad de Valparaíso, General Cruz 222, Valparaíso, Chile}
\email{rodrigo.meneses@uv.cl}

\author[0000-0002-4502-6330,gname=Michaela,sname=Kraus]{Michaela Kraus}
\affiliation{Astronomical Institute, Czech Academy of Science, Fri\v{c}ova 298, 251 65 Ond\v{r}ejov, Czech Republic}
\email{michaela.kraus@asu.cas.cz}

\begin{abstract}

Magnetic null points define the topology of magnetic fields in stellar coronae, planetary magnetospheres, and astrospheres, and are commonly associated with magnetic reconnection. We investigate the local structure of stationary three-dimensional (3D) resistive, incompressible magnetohydrodynamic (MHD) flows in the vicinity of an isolated magnetic null and an isolated velocity stagnation point, assuming that the two nulls are colocated. The governing MHD equations are solved self-consistently using a first-order local expansion of the magnetic and velocity fields about the null. For any given magnetic-null configuration, the equations admit a two-parameter family of stationary solutions, comprising multiple solution branches with distinct velocity topologies. Along these families, the velocity-null polarity can change at bifurcation points where the flow becomes effectively 2.5D. A single magnetic null can therefore support positive-radial, negative-radial, and spiral stagnation flows, with qualitatively different inflow, outflow, and rotational properties. The associated pressure and resistivity distributions are likewise determined self-consistently and vary across the solution family, demonstrating that they are coupled to the flow topology rather than prescribed independently. The generic solutions do not require the velocity and magnetic separatrices to coincide, allowing plasma to cross the magnetic fan separatrix and providing a local reconnective flow. These results demonstrate that, in the presence of nonuniform resistivity, the magnetic topology alone does not uniquely determine the associated stationary flow, pressure, or resistivity structure around a 3D magnetic null, revealing the nonlinear feedback between all macroscopic variables and resistivity. The constant-resistivity field-aligned and spiral field-crossing solutions identified previously are recovered as special cases.

\end{abstract}

\keywords{\uat{Magnetohydrodynamics}{1964} ---  \uat{Magnetic fields}{994} --- \uat{Stellar-interstellar interactions}{1576} --- \uat{Astrospheres}{107} --- \uat{Solar magnetic reconnection}{1504}}


\section{Introduction}
\label{sec:introduction}

Magnetic null points are common features of cosmic magnetic fields and frequently associated with magnetic reconnection and eruptive activity. Early observational evidence for an isolated three-dimensional (3D) null point in the solar corona was reported by \cite{filippov1999}. Subsequent observations have identified coronal null points in association with circular-ribbon flares \citep{masson2009}, eruptive flares and coronal mass ejections \citep{sun2016}, spectroscopically resolved flare reconnection \citep{yang2020}, and persistent magnetic reconnection revealed by ultrahigh-resolution observations \citep{cheng2023}. Complementary 3D magnetohydrodynamical (MHD) simulations reproduce these magnetic configurations and demonstrate that null-point reconnection naturally drives circular-ribbon flares, recurrent coronal jets through the release of twist accumulated in the magnetic field, and eruptive events triggered by magnetic flux emergence \citep{pariat2009,pariat2010,morenoinsertis2013}. Statistical studies further suggest that magnetic null points are associated with a significant fraction of X-class flares, highlighting their importance in solar eruptive activity \citep{edgar2024}.
Isolated magnetic null points have been observed throughout the terrestrial magnetosphere across a wide range of plasma environments \citep{guo2022}. Radial nulls have been identified in the bow shock transition region, ion foreshock, and dayside magnetopause, where they are associated with magnetic reconnection and electron energization within electron diffusion regions \citep{chen2019,jiang2021,dunlop2009,fu2019}. In the turbulent magnetosheath, isolated and spiral nulls organize the 3D reconnection geometry and local current structure \citep{wendel2013,chen2018,tian2025}, while in the magnetotail isolated nulls have been associated with particle acceleration and wave activity within reconnection diffusion regions \citep{xiao2006}.

Analytical and numerical models of the heliosphere
and astrospheres likewise indicate that isolated magnetic null points may govern the large-scale topology of stellar-wind interaction regions. The heliopause (or astropause) is commonly described as the separatrix separating the stellar wind from the surrounding interstellar medium, with a stagnation point located near its nose \citep{belov2010,herbst2022}. In a topological interpretation of this configuration, the stagnation point has been identified with a magnetic null, implying that the magnetic and hydrodynamic separatrices coincide locally and that the distribution of magnetic nulls determines the geometry of the heliopause and astropauses in general \citep{nickeler2006,2014ASTRP...1...51N,nickeler2008}.

In the vicinity of an isolated 3D null, the magnetic field is organized into a pair of spine field lines and a fan separatrix surface, which divide the surrounding volume into distinct regions of magnetic connectivity \citep{lau1990,priesttitov1996}. Depending on the local eigenstructure of the magnetic field, nulls may be classified as radial or spiral, with electric currents modifying the geometry of the spine--fan configuration while preserving its underlying topological structure \citep{parnell1996}. Together, the spine, fan, and associated separatrix surfaces constitute the magnetic skeleton that governs the connectivity of magnetic field lines and provides the topological framework for 3D magnetic reconnection \citep{longcope2005,pontin2022}.

Analytical studies of 2D magnetic reconnection span the classical Sweet--Parker and Petschek models of resistive diffusion \citep{sweet1958,petschek1964}, incompressible MHD analyses relating X-type magnetic fields to reconnection and stagnation-point flows to magnetic annihilation \citep{priest1975}, exact stationary solutions for reconnective annihilation and X-point reconnection \citep{craig1991,craig1995}, and more recent topological formulations that couple the magnetic field, plasma flow, and spatially varying resistivity \citep{nickeler2012,nickeler2019}. 

In 3D, the analytical description of magnetic reconnection progressed through a sequence of complementary approaches. \citet{priesttitov1996} established the topological distinction between spine and fan reconnection and demonstrated that the corresponding ideal plasma flows become singular in the vicinity of 
3D nulls, motivating resistive MHD descriptions of steady reconnection. Exact stationary solutions of the incompressible resistive MHD equations were subsequently derived for generic
3D nulls, identifying field-aligned stagnation flows and, for sufficiently strong spine-parallel currents, spiral field-crossing solutions. Despite the presence of a finite electric field at the null, these linear, constant resistivity solutions remain nonreconnective in nature, with no plasma transport across the spine or fan of the magnetic field \citep{titov2000}. Rather than solving the full resistive MHD equations, a kinematic formulation was developed in which the magnetic field and localized nonideal region are prescribed while the momentum equation is omitted \citep{hornigpriest2003}. Within this framework, spine- and fan-aligned current configurations were shown to produce distinct reconnection regimes characterized by rotational slippage and magnetic flux transfer, respectively \citep{pontin2004,pontin2005}. These models were later generalized to asymmetric magnetic nulls and to spatially localized current layers while retaining the prescribed magnetic geometry and kinematic formulation \citep{alhachami2010,pontin2011}. Numerical solutions of the full resistive MHD equations subsequently reproduced these reconnection regimes, demonstrating current accumulation along the spine or fan under rotational driving, spine--fan reconnection under shear perturbations, and the influence of magnetic asymmetry, boundary driving, and null-point collapse on current-sheet formation and reconnection dynamics \citep{pontingalsgaard2007,priestpontin2009,galsgaardpontin2011,wyper2012,thurgood2018}. 

Previous analytical and numerical studies have established the topological structure and reconnection regimes associated with
3D magnetic nulls. The families of stationary resistive MHD solutions that are locally admissible around a generic null, however, remain comparatively unexplored. In the present work, we investigate the local structure of such solutions in the vicinity of an isolated 3D magnetic null. Rather than prescribing a particular reconnection geometry, we derive the stationary solutions permitted by the governing equations and investigate the relationships among the local MHD fields implied by them. The main objective is to investigate the relationship between the topology of the magnetic null and the surrounding plasma flow, provided by the Poincar\'{e}--Lyapunov classification, which is based on the eigenstructure of both of the involved vector fields \citep[see, for example,][]{nickeler2019}. We obtain different families of admissible stationary flows, as well as the occurrence of topological bifurcations in the analysis. 

The paper is organized as follows. Section~\ref{sec:equations} introduces the assumptions and governing MHD equations, together with the local Taylor expansion of the fields and the resulting stationary equations. Section~\ref{sec:methods} describes the solution procedure used to obtain the stationary solution families. A representative example is then used to investigate the parametric dependence of the different configurations of the local plasma-flow dynamics. In Section~\ref{sec:results}, we characterize the resulting solutions for a fiducial magnetic null, including their flow, pressure, and resistivity structures. Section~\ref{sec:discussion} discusses the astrophysical implications of the solutions for coronal jets and astrospheres, examines their reconnective properties, and considers the role of nonconstant resistivity. Finally, Section~\ref{sec:conclusions} summarizes our main results and conclusions.

\section{Assumptions and Basic Equations}
\label{sec:equations}

We seek local stationary solutions of the resistive MHD equations in the vicinity of an isolated 3D magnetic null point. The plasma dynamics are assumed to be highly subsonic, and the flow is treated as incompressible. Dissipative effects other than Ohmic diffusion, such as viscosity and thermal conduction, are neglected. Under these assumptions, the governing dimensionless resistive MHD equations comprise the momentum equation, generalized Ohm's law, the incompressibility condition, Ampère's law (neglecting the displacement current), Faraday's law, and Gauss's law for magnetism \citep{goedbloed2004}:
\begin{align}
	\frac{\partial \textbf{v}}{\partial t} + (\textbf{v} \cdot \boldsymbol\nabla) \textbf{v} &= -\boldsymbol\nabla p + \textbf{j} \times \textbf{B}, \label{eq:momentum}\\
	\textbf{E} + \textbf{v} \times \textbf{B} &= \eta \textbf{j}, \label{eq:ohm}\\
	\boldsymbol\nabla \cdot \textbf{v} &= 0, \label{eq:incompressibility} \\
	\boldsymbol\nabla \times \textbf{B} &=  \textbf{j}, \label{eq:ampere}\\
	\boldsymbol\nabla \times \textbf{E} &= -\frac{\partial \textbf{B}}{\partial t}, \label{eq:faraday} \\
	\boldsymbol\nabla \cdot \textbf{B} &= 0, \label{eq:gauss}
\end{align}
Here, $\mathbf{v}$ denotes the plasma velocity, $p$ the plasma pressure, $\mathbf{B}$ the magnetic field, $\mathbf{E}$ the electric field, $\mathbf{j}$ the electric current density, and $\eta$ the electrical resistivity.
The magnetic field, the velocity and the pressure are normalized in terms of a characteristic magnetic field $B_{0}$, the Alfv\'{e}n speed $v_{\text{A}0}$, and the ram pressure $\tfrac{1}{2}\rho_{0} v_{\text{A}0}^{2}$, respectively.
We assume a uniform mass density $\rho_{0}$, which is absorbed into the normalization of the dimensionless pressure and magnetic variables. The length and time are normalized to some characteristic length scale $l_{0}$ and $l_{0}/v_{\text{A}0}$. Consequently, the resistivity is normalized to $\tilde{\eta} = v_{\text{A}0}^{2} l_{0} \mu_{0},$ where $\mu_{0}$ is the permeability of free space. 

This analysis focuses on the particular configuration where the magnetic null and the velocity stagnation point coincide at the origin of the coordinate system. We denote the locations of the magnetic null and the stagnation point by $\mathbf{r}_{\text{null}}$ and $\mathbf{r}_{\text{stag}}$, respectively, such that $\mathbf{B}(\mathbf{r}_{\text{null}}) = \mathbf{v}(\mathbf{r}_{\text{stag}}) = 0$, and study the relationship between the topology induced by the magnetic null and the flow associated with the stagnation point. Defining
\begin{equation}
    \delta = |\mathbf{r}_{\text{null}} - \mathbf{r}_{\text{stag}}|,
\end{equation}
this parameter also affects the relationship between the magnetic and kinematic geometries. Therefore, considering $\delta = 0 $ not only removes a free parameter from the analysis, but also directly links the Jacobian matrices of the magnetic and velocity fields. This setting may also enhance the probability of obtaining structurally stable configurations.

Following the local formulation of \citet{titov2000}, the MHD variables are assumed to be sufficiently smooth and are Taylor expanded about the origin. Retaining only the leading-order terms, the velocity and magnetic fields are represented by linear functions of position, while the pressure is retained to second order to satisfy the momentum equation:
\begin{align}
	\textbf{v} &= \mathcal{V}\cdot \textbf{r}, \label{eq:vexp}\\
	\textbf{B} &= \mathcal{B} \cdot \textbf{r}, \label{eq:bexp}\\
	p &= \frac{1}{2}\, \textbf{r}^{\mathrm T}\cdot\tilde{\mathcal{P}}\cdot\textbf{r} + p_{0}. \label{eq:pexp}
\end{align}
Here, $\mathbf{r}$ is the position vector, the superscript $\mathrm{T}$ represents the transpose, $\mathcal{V}(t)$ and $\mathcal{B}(t)$ are the Jacobian matrices of the velocity and magnetic fields, and
$\mathcal{P}(t)$ is the half of the Hessian matrix of the pressure. Their components are
$\mathcal{V}_{ij}=\partial v_i/\partial x_j$,
$\mathcal{B}_{ij}=\partial B_i/\partial x_j$, and
$\tilde{\mathcal{P}}_{ij}=\partial^2p/\partial x_i\partial x_j$.

From Ohm's law, Equation~\eqref{eq:ohm}, the spatial dependence of the resistivity can be inferred from the orders of the fields in the spatial coordinates. The magnetic and velocity fields are linear in $x$, $y$ and $z$, while the electric field is quadratic. Ampère's Law, Equation~\eqref{eq:ampere}, implies that the current density is constant for a linear magnetic-field configuration, and hence, the resistivity obtained from Ohm's law can contain terms of at most second order. The general form can thus be written as
\begin{equation}
    \eta = \eta_0 + \boldsymbol{\eta}_1\cdot\mathbf r
    +\frac{1}{2}\, \mathbf r^{\mathsf T} \cdot \tilde{\mathcal{R}}\cdot\mathbf r,
    \label{eq:rexp}
\end{equation}
where $\mathbf r=(x,y,z)^{\mathsf T}$, $\boldsymbol{\eta}_1$ is a vector
containing the coefficients of the linear terms, $\eta_{0}$ is the resistivity at the null, and $\tilde{\mathcal{R}}$ is the
resistivity Hessian. For notational convenience, we define the matrices $\mathcal{P} = 2 \tilde{\mathcal{P}}$ and $\mathcal{R} = 2 \tilde{\mathcal{R}}$. As for any Hessian of a scalar field, $\mathcal{P}$ and $\mathcal{R}$ are symmetric, and therefore have only six independent components.

\subsection{Stationary Flows}

We now restrict our attention to stationary solutions, for which all time derivatives vanish. Upon substituting Equations~(\ref{eq:vexp})--(\ref{eq:pexp}) into Equation~(\ref{eq:momentum}), the antisymmetric part becomes independent of the pressure owing to the symmetry of the pressure Hessian, $\mathcal{P}^{\mathrm T}=\mathcal{P}$. Together with Equations~(\ref{eq:incompressibility}), (\ref{eq:faraday}), and (\ref{eq:gauss}), the local stationary equations reduce to
\begin{align}
	\mathcal{V}^{2} - \left( \mathcal{V}^{\mathrm T} \right)^{2} &= \mathcal{B}^{2} - \left( \mathcal{B}^{\mathrm T} \right)^{2}, \label{eq:st_momentum}\\
	\text{tr}\left( \mathcal{V} \right) &= 0, \label{eq:st_incompressible}\\
	\boldsymbol\nabla \times \textbf{E} &= 0, \label{eq:st_faraday}\\
	\text{tr}\left( \mathcal{B} \right) &= 0. \label{eq:st_gauss}
\end{align}
The symmetric part of the momentum equation determines the pressure Hessian,
\begin{align}
	\mathcal{P} = \mathcal{B}^{2} - \mathcal{B}^{\mathrm T}\mathcal{B} - \mathcal{V}^{2}.
    \label{eq:pHess}
\end{align}
After an appropriate rotation and scaling of the coordinate system, the magnetic field Jacobian assumes the form \citep{parnell1996},
\begin{equation}
	\mathcal{B} =
	\begin{pmatrix}
		1 & \left(q - j_{\parallel} \right)/2 & 0 \\
		\left(q + j_{\parallel} \right)/2 & b & 0 \\
		0 & j_{\perp} & -\left(b + 1 \right) 
	\end{pmatrix}.
\end{equation}
The symmetric part of $\mathcal{B}$ describes the potential component of the magnetic field and is characterized by the parameters $b$ and $q$, whereas the antisymmetric part is determined by the electric current density,
\begin{equation}
	\textbf{j} = \left(
	\begin{matrix}
		j_{\perp} \\
		0 \\
		j_{\parallel}
	\end{matrix}
	\right),
\end{equation}
where the subscripts $\parallel$ and $\perp$ refer to components parallel and perpendicular to the spine (the $z$-axis), respectively. Throughout this work we assume $j_\parallel\neq0$ and $j_\perp\neq0$. To ensure that the spine coincides with the eigenvector $\mathbf{x}_3=(0,0,1)^{\mathrm T}$, associated with the eigenvalue $-(b+1)$, the parameters are restricted according to \citet{parnell1996},
\begin{equation}
\begin{aligned}
b &> -1,\\
j_{\parallel}^{2} &> q^{2}-4b.
\end{aligned}
\end{equation}

\section{Solving the Stationary Equations}
\label{sec:methods}

The stationary resistive MHD equations are solved by successively determining the electrostatic potential, resistivity, and admissible velocity Jacobians. Here we summarize the main steps of the derivation, while the detailed calculations are given in Appendix~\ref{app:derivation}. For stationary solutions, Equation~\eqref{eq:st_faraday} implies that the electric field is irrotational and may therefore be written as
\begin{equation}
    \mathbf{E}=-\boldsymbol\nabla\phi,
\end{equation}
where $\phi$ is the electrostatic potential. Substituting this expression into Ohm's law, Equation~\eqref{eq:ohm}, gives the component equations
\begin{subequations}\label{eq:ohms}
\begin{align}
-\frac{\partial\phi}{\partial x}+v_yB_z-v_zB_y
&=\eta\,j_{\perp},\label{eq:ohms_a}\\
-\frac{\partial\phi}{\partial y}+v_zB_x-v_xB_z
&=0,\label{eq:ohms_b}\\
-\frac{\partial\phi}{\partial z}+v_xB_y-v_yB_x
&=\eta\,j_{\parallel},\label{eq:ohms_c}
\end{align}
\end{subequations}
\noindent where $ v_{x} $, $ v_{y} $, $ v_{z} $ and $ B_{x} $, $ B_{y} $, $ B_{z} $ are the $x-$, $y-$ and $z-$components of the velocity and the magnetic field respectively. Since the present analysis is restricted to the local neighborhood of the magnetic null, the velocity and magnetic fields are represented to first order by Equations~\eqref{eq:vexp} and \eqref{eq:bexp}. Consequently, Equations~\eqref{eq:ohms} require the electrostatic potential $\phi$ to be a cubic polynomial and the resistivity $\eta$ to be at most quadratic in the spatial coordinates.

Integrating Equation~\eqref{eq:ohms_b} with respect to $y$ yields
\begin{equation}
\label{eq:phi_sol}
    \phi(x,y,z)=F(x,y,z)+g(x,z),
\end{equation}
where $F(x,y,z)$ denotes the polynomial contribution arising from the linear velocity and magnetic fields, while $g(x,z)$ is an arbitrary function of integration. Eliminating the resistivity by multiplying Equation~\eqref{eq:ohms_a} by $j_{\parallel}$, Equation~\eqref{eq:ohms_c} by $j_{\perp}$, and subtracting the resulting equations yields a hyperbolic first-order partial differential equation (PDE) for $g(x,z)$ of the form
\begin{equation}
\label{eq:g_PDE}
    j_{\parallel}\frac{\partial g(x,z)}{\partial x} - j_{\perp}\frac{\partial g(x,z)}{\partial z}=S(x,z),
\end{equation}
where $S$ is a polynomial source term. This equation is readily solved by the method of characteristics, giving
\begin{equation}
\label{eq:g_sol}
    g(x,z)=G(x,z)+f(j_{\perp}x+j_{\parallel}z),
\end{equation}
where $G(x,z)$ is a polynomial particular solution and $f$ is an arbitrary function of the argument $j_{\perp}x+j_{\parallel}z$. To obtain solutions that are locally analytic in the vicinity of the magnetic null, the arbitrary function $f$ is Taylor expanded about the origin. Since the electrostatic potential is required to be cubic in the spatial coordinates, consistent with the first-order expansions of the magnetic and velocity fields, the expansion is truncated after the third-order term,
\begin{align}
\label{eq:f_expansion}
f\left(j_{\perp}x+j_{\parallel}z\right)
& \nonumber=
c_{1}\left(j_{\perp}x+j_{\parallel}z\right)^{3}
+c_{2}\left(j_{\perp}x+j_{\parallel}z\right)^{2} \\ 
&+c_{3}\left(j_{\perp}x+j_{\parallel}z\right)
+c_{4},
\end{align}
where $c_1\ldots c_4$ are arbitrary constants.

Finally, substituting the electrostatic potential into Equations~\eqref{eq:ohms_a} and \eqref{eq:ohms_c} yields two expressions for the resistivity $\eta(x,y,z)$. Since both expressions represent the same scalar field, they must coincide identically, giving rise to three linear constraints on the elements of the velocity Jacobian. The momentum equation contributes three additional constraints that are quadratic in the velocity and magnetic field parameters, while the incompressibility condition provides a fourth linear constraint. Together, these constitute seven independent constraints on the nine components of the velocity Jacobian. The seven constraints have full rank, so that the admissible velocity Jacobians form a locally two-dimensional solution set. The velocity Jacobian may therefore be parametrized by two independent variables, which we choose as $V_{11}$ and $V_{13}$. 
With this choice, the three momentum equations reduce to a single polynomial equation of degree six in the unknown $V_{12}$, whose coefficients are polynomials in the free parameters $V_{11}$ and $V_{13}$. The coefficient of $V_{12}^{6}$ is a nonzero constant, independent of $V_{11}$ and $V_{13}$. Consequently, the polynomial remains of degree six throughout the parameter space, yielding six roots (counting multiplicities) at every point. In particular, no root can escape to infinity as the parameters are varied. Since none of these roots admits a closed-form expression, they must be determined numerically on a grid spanning the parameter space $\left(V_{11},V_{13}\right)$. Detailed expressions for the electrostatic potential,  resistivity, and the resulting constraints are provided in Appendix~\ref{app:derivation}.

The present formulation recovers the constant-resistivity solutions of \citet{titov2000} as a special case. Specifically, setting $c_{1}=c_{2}=0$ in Equation~\eqref{eq:f_expansion}, together with $V_{11}=1$ and $V_{13}=0$, yields $\eta=-c_{3}$ and the field-aligned solutions,
\begin{equation}
    \mathcal{V}=\pm\mathcal{B}.
\end{equation}

\section{Properties of the Solutions}
\label{sec:results}

We demonstrate the topological properties of the MHD fields through an illustrative example. For this, we consider the magnetic field 
\begin{align}
\mathbf{B}(x, y, z) = (x + 0.1 y,\ 0.2 x + 0.5 y,\ 0.1 y - 1.5 z),
\label{exB1}
\end{align}
with the corresponding parameter values
\begin{align}
b = 0.5,\quad q = 0.3,\quad j_{\parallel} = 0.1,\quad \text{and} \quad
j_{\perp} = 0.1,
\label{exB1params}
\end{align}
which corresponds to a positive improper radial null. Here, \emph{positive} indicates that the two eigenvalues associated with the fan are positive, while the spine eigenvalue is negative, whereas \emph{improper radial} denotes that the fan eigenvalues are real and unequal.
To characterize the families of stationary solutions, we fix $V_{13} = 0$ and examine their dependence on the remaining free parameter $V_{11}$, over the range $ \left[ -2, 2\right] $. Beyond this interval, the solutions exhibit no qualitatively new behavior for the prescribed magnetic-field configuration.
The corresponding reduction of the velocity constraints is given in Appendix~\ref{app:example}.

\begin{figure*}[t!]
    \centering
    \begin{subfigure}{0.48\linewidth}
        \centering
        \includegraphics[width=0.96\linewidth]{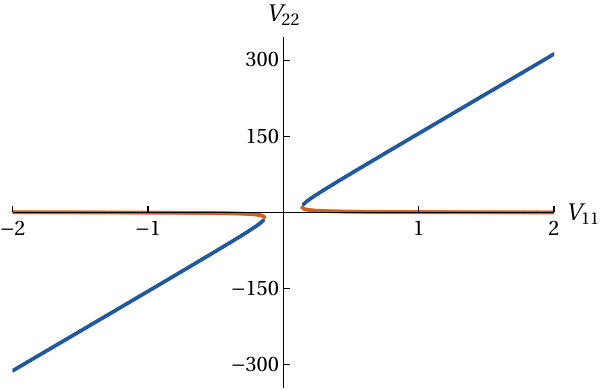}
        \caption{}
        \label{fig:V22comp}
    \end{subfigure}%
    \begin{subfigure}{0.48\linewidth}
        \centering
        \includegraphics[width=0.96\linewidth]{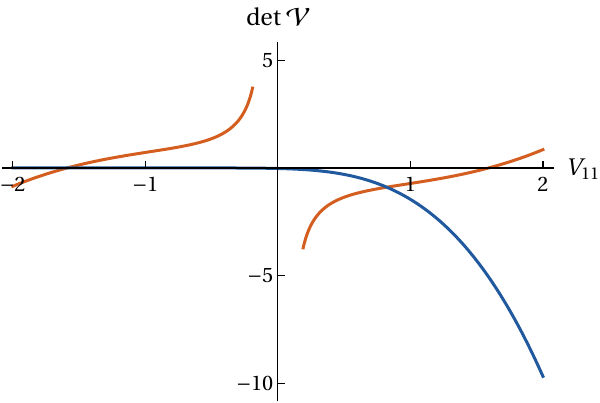}
        \caption{}
        \label{fig:detVcomp}
    \end{subfigure}
    \caption{\textbf{Left:} $V_{22}$ as a function of $V_{11}$ for branch $k=1$ (orange) and $k=4$ (blue). The gap indicates the interval of $V_{11}$ over which the solutions become complex.
    \textbf{Right:} $\det\mathcal{V}$ as a function of $V_{11}$ for branches $k=1$ (orange) and $k=2$ (blue).}
\end{figure*}

\subsection{Structure of Flows}

We denote the stationary solutions for the velocity Jacobian by $\mathcal{V}^{(k)} $, where the superscript $k$ labels the different solution branches corresponding to the distinct roots of the governing polynomial. Hereafter, we refer to each solution simply as a \emph{branch}. Within the prescribed interval $ V_{11} \in \left[ -2,2 \right],  $ two of the six branches remain complex throughout the entire domain and are therefore excluded from further consideration. Consequently, only the four real branches, labeled by $k = 1, \ldots, 4$ are analyzed. These branches satisfy the symmetry relations
\begin{equation}
\mathcal{V}^{(k)}(V_{11})=
\begin{cases}
-\mathcal{V}^{(k)}(-V_{11}), & k\in\{1,4\}, \\
-\mathcal{V}^{(k+1)}(-V_{11}), & k=2,
\end{cases}
\label{eq:branchsym}
\end{equation}
while the branch with $k=3$ is obtained directly from the second relation. The dependence of each velocity Jacobian element and the determinant on the free parameter $V_{11}$ for each branch is presented in Figure~\ref{fig:app_solutions}.

Several structural properties further restrict the set of branches under consideration. Branches $k=1$ and $k=4$ merge at $V_{11} \sim \pm 0.1386 $ and continue as a complex-conjugate pair, so no real solution exists on the interval $V_{11} \in (-0.1386,0.1386) $ (Figure~\ref{fig:V22comp}). 
On branch $k=4$ the components $V_{12}$ and $V_{21}$ nearly coincide, and since $ V_{22} \propto 1/(V_{12} - V_{21}) $ the component $V_{22}$ diverges accordingly, reaching $|V_{22}| \sim 300$ for $|V_{11}| \sim 2$ (Figure~\ref{fig:V22comp}).  
Across the range studied, $\max|V_{ij}|$ on this branch never falls below $12.6$, whereas it does not exceed $9.5$ on branch $k=1$ and $5.5$ on branches $k=2,3$. The magnetic Jacobian is normalized such that $\max\,|\mathcal{B}_{ij}|=1$, so that $\mathcal{B}$ sets the reference scale against which the velocity gradients are measured. A solution with $\max\,|V_{ij}|\gtrsim 10^{2}$ therefore describes a flow whose shear exceeds the magnetic-field gradient by more than two orders of magnitude. Hence, the radius within which the first-order velocity term dominates is correspondingly smaller than that within which the adopted first-order local approximation of the magnetic field is valid. Consequently, within the common domain of validity, the neglected higher-order velocity terms cannot be assumed to be small. We therefore regard branch $k=4$ as physically inconsistent and exclude it from the subsequent analysis. Branches $k=2$ and $k=3$ are related by the symmetry of Equation~\eqref{eq:branchsym}, that is, by reflection in $V_{11}$ combined with reversal of the flow direction, which exchanges the polarity of the null. The two branches therefore carry the same information, and in what follows we analyze branch $k=2$ over the full range $V_{11}\in[-2,2]$.

On this slice, where $V_{13}=0$, the linear constraint gives $V_{23}=0$, and $\mathcal{V}^{(k)}$ is block lower-triangular. Its characteristic polynomial therefore factorizes,
\begin{align}
F& \nonumber^{(k)}(\lambda) = \det\!\left(\mathcal{V}^{(k)} - \lambda I\right) \\ &
= \left(V_{33}-\lambda\right)
  \left[\lambda^{2} - \left(V_{11}+V_{22}\right)\lambda
        + \left(V_{11}V_{22}-V_{12}V_{21}\right)\right],
\label{eq:charpoly}
\end{align}
where $\lambda$ denotes the eigenvalue and $I$ the identity matrix. One eigenvalue is thus $V_{33}$, and the remaining two are those of the upper-left $2\times 2$ block, which we designate with $M$. The incompressibility condition Equation~\eqref{eq:st_incompressible} fixes $V_{33}=-\operatorname{tr}M$.
The eigenvalues are therefore elementary functions of the Jacobian components,
although the components themselves are available only numerically as functions
of $V_{11}$.

Setting $\lambda=0$ in Equation~\eqref{eq:charpoly} gives
\begin{align}
\det\!\left(\mathcal{V}^{(k)}\right)
= -\left(V_{11}+V_{22}\right)\left(V_{11}V_{22}-V_{12}V_{21}\right) = 0,
\label{bfc1}
\end{align}
which identifies the values of $V_{11}$ at which one eigenvalue vanishes and the
null becomes degenerate. These are the points at which the polarity of the
velocity null changes sign, and they are shown in Figure~\ref{fig:detVcomp}.

As demonstrated below, these points partition the parameter range into distinct intervals, each characterized by a qualitatively different structure of the local velocity field. We therefore refer to them as \textit{bifurcation points}. In the
full $(V_{11},V_{13})$ parameter space they are the traces of bifurcation curves on the slice $V_{13}=0$.

\begin{figure}[htbp]
  \centering
  \includegraphics[width=0.47\textwidth]{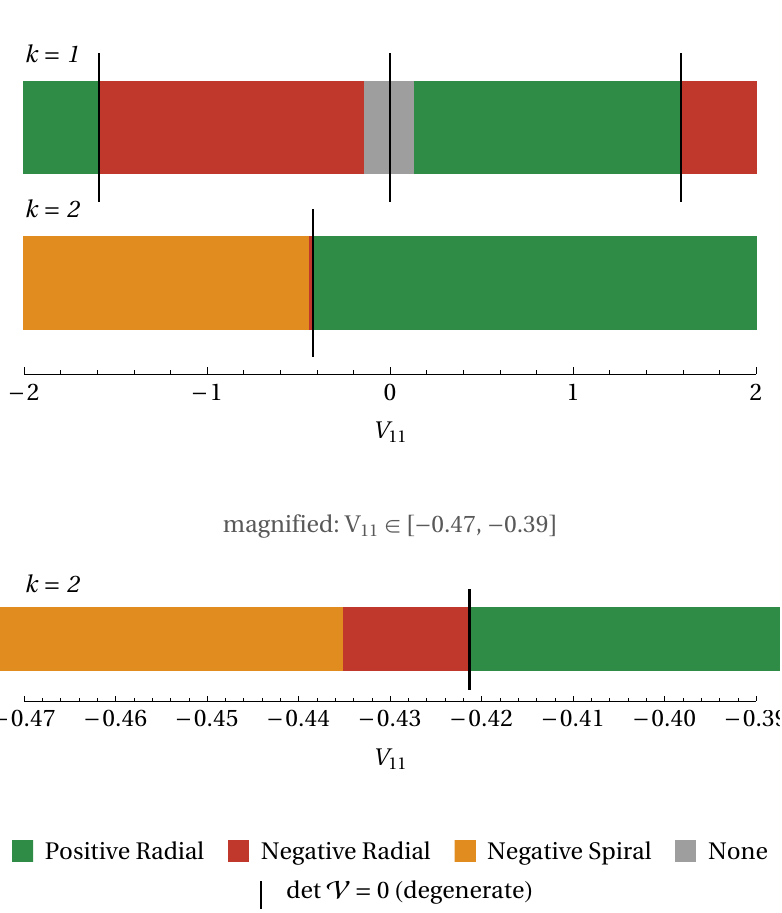}
  \caption{Classification of the velocity null along the slice $V_{13}=0$ for the solution branches $k=1$ (top) and $k=2$ (bottom). Colors denote the local flow topology according to the legend below. Vertical black lines mark bifurcation points. Grey regions denote intervals where the branch has no real solutions. The lower panel magnifies $V_{11}\in[-0.47,-0.39]$ for branch $k=2$, resolving the narrow negative-radial interval of width $0.0117$ between the bifurcation point and the radial-to-spiral transition.}
  
  \label{fig:ex1Bfc}
\end{figure}

We evaluate $\det\mathcal{V}^{(k)}$ along each branch and locate its zeros. Crossing such a zero, one eigenvalue changes sign, causing the velocity null to reverse its polarity. These are genuine topological transitions of the local flow. A second type of boundary occurs where $\det M = \left(V_{11}-V_{22}\right)^{2} + 4\,V_{12}V_{21}$ changes sign and the eigenvalue pair of $M$ passes from real to complex. There the null changes from radial to spiral in the standard classification, although the two flows remain topologically conjugate. The resulting classification of
each branch is shown in Figure~\ref{fig:ex1Bfc}.

On branch $k=1$, the solution is antisymmetric about $V_{11}=0$, as given by Equation~\eqref{eq:branchsym}. For $V_{11}>0$, the null is positive radial for $0.1358<V_{11}<1.5898$ and negative radial for $V_{11}>1.5898$, with $\det\mathcal{V}^{(1)}=0$ at $V_{11}\simeq1.5898$ marking the bifurcation between the two configurations. No real solutions exist for $0<V_{11}<0.1358$, represented by
the gray region in Figure~\ref{fig:ex1Bfc}. The corresponding behavior for $V_{11}<0$ follows from the antisymmetry, with the null topology reversed and with the transition from positive to negative radial occurring through $\det\mathcal{V}^{(1)}=0$ at $V_{11}=0$. The eigenvalues remain real wherever solutions exist, so no spiral configuration occurs.
Branch $k=2$ exhibits three distinct configurations as $V_{11}$ is varied. It is a negative spiral null for $V_{11}\lesssim-0.433$, where $\det M<0$. At $V_{11}\sim-0.433$, $\det M$ changes sign and the complex-conjugate eigenvalue pair becomes real, yielding a negative radial null. At $V_{11}\sim-0.4213$, only $\sim0.0117$ farther along the branch, the determinant vanishes and the null changes to a positive radial configuration, which persists over the remainder of the range. The narrow negative-radial interval between these two transitions is resolved in the magnified panel.

\begin{figure*}[htbp]
\label{fig:ex1_VFlow}
  \centering
  
  \begin{subfigure}[b]{0.33\textwidth}
    \includegraphics[width=\linewidth]{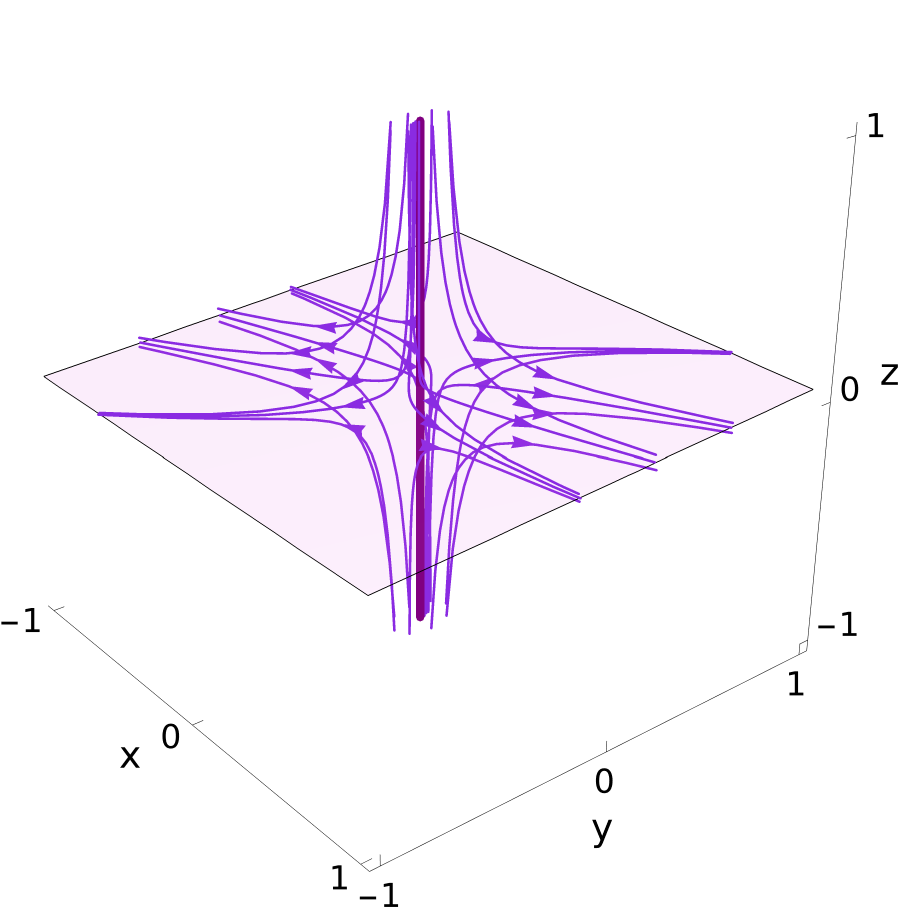}
    \caption{}
    \label{fig:ex1_B}
  \end{subfigure}%
   \hfill
  \begin{subfigure}[b]{0.33\textwidth}
    \includegraphics[width=\linewidth]{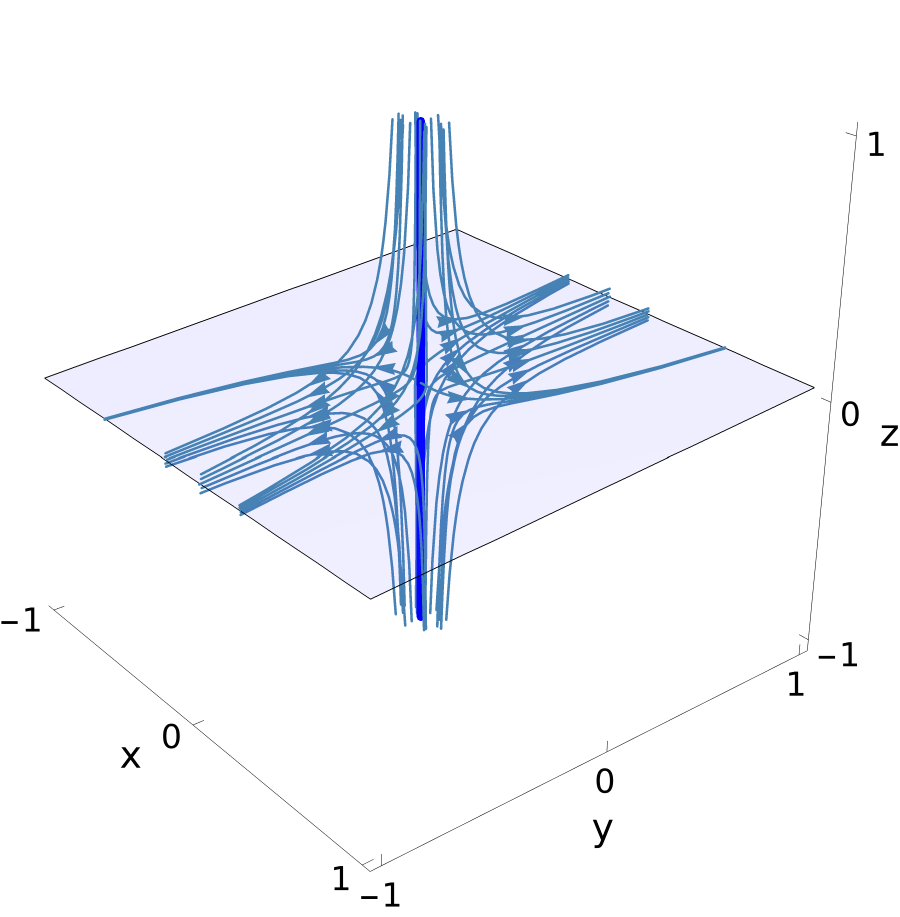}
    \caption{$k = 1;\, V_{11} = 0.5$}
    \label{fig:k1PRN}
  \end{subfigure}%
  \hfill
  \begin{subfigure}[b]{0.33\textwidth}
    \includegraphics[width=\linewidth]{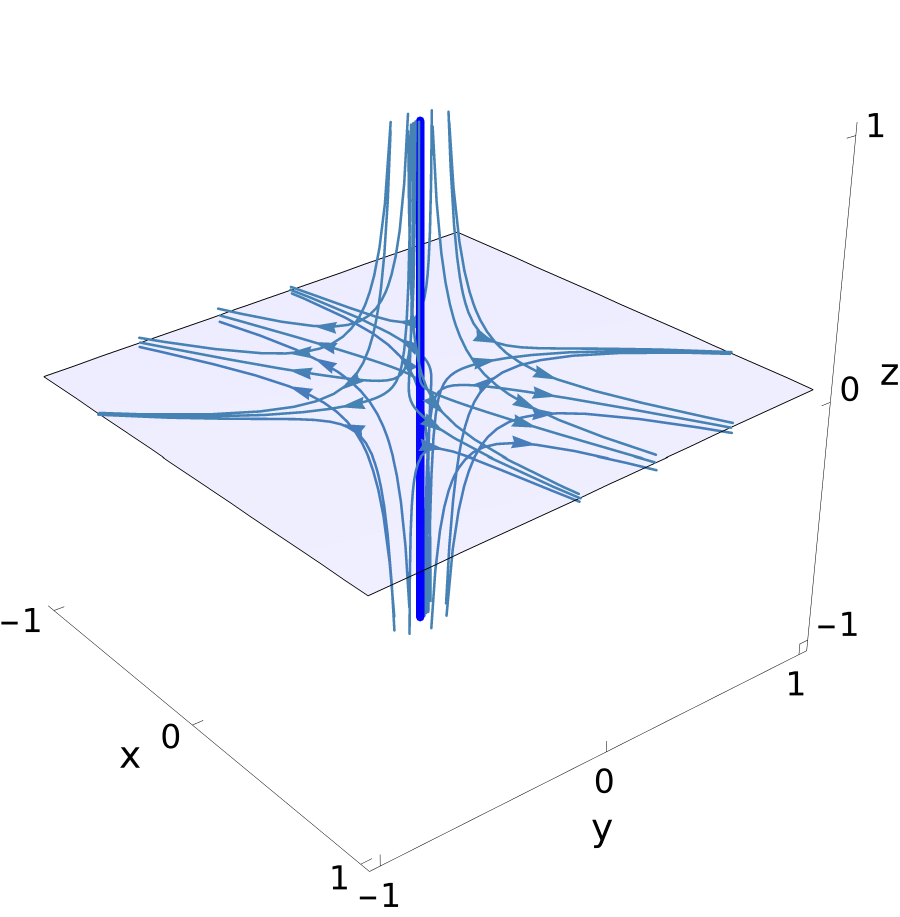}
    \caption{$k = 1;\, V_{11} = 1.0$}
    \label{fig:k1FA}
  \end{subfigure}%

  \vspace{1ex}

  \begin{subfigure}[b]{0.33\textwidth}
    \includegraphics[width=\linewidth]{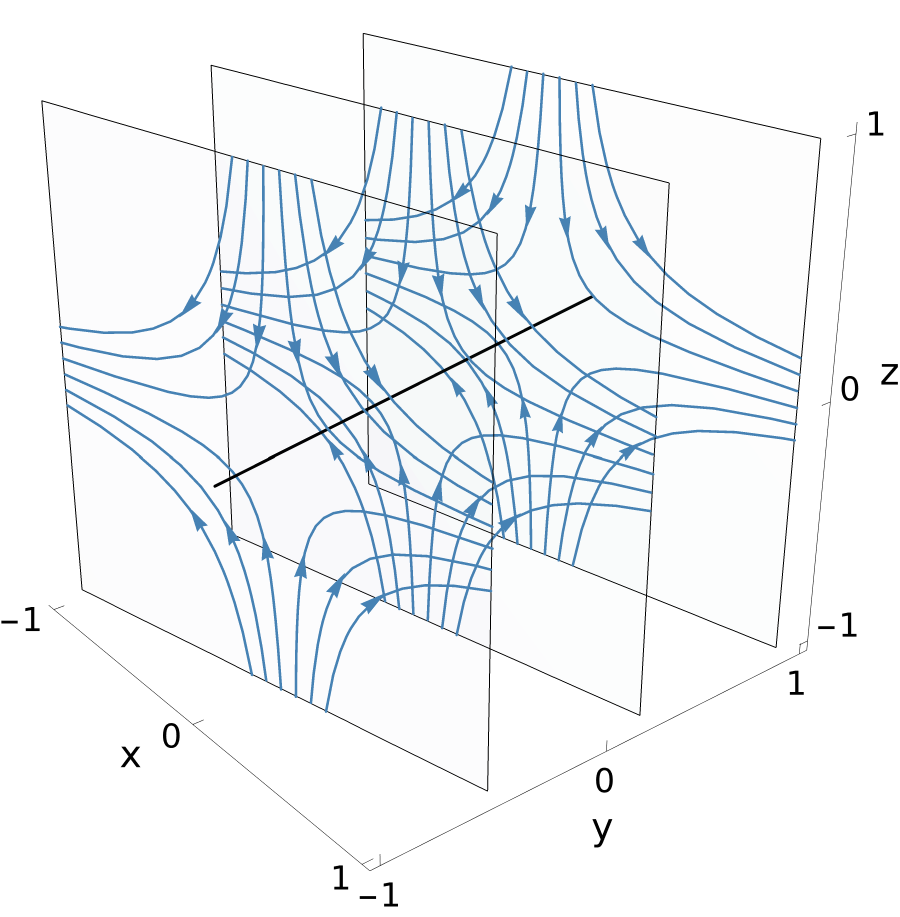}
    \caption{$k = 1;\, V_{11} = 1.5898$}
    \label{fig:k1Deg}
  \end{subfigure}
  \hfill
  \begin{subfigure}[b]{0.33\textwidth}
    \includegraphics[width=\linewidth]{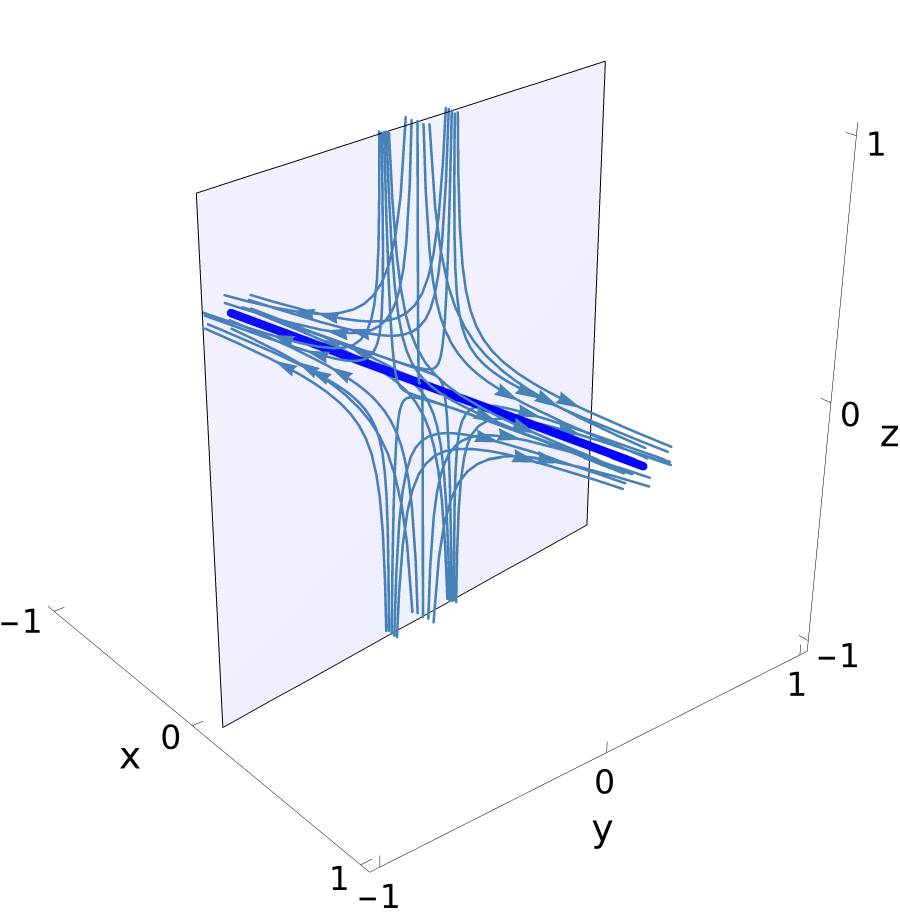}
    \caption{$k = 1;\, V_{11} = 2.0$}
    \label{fig:k1NRN}
  \end{subfigure}%
  \hfill
  \begin{subfigure}[b]{0.33\textwidth}
    \includegraphics[width=\linewidth]{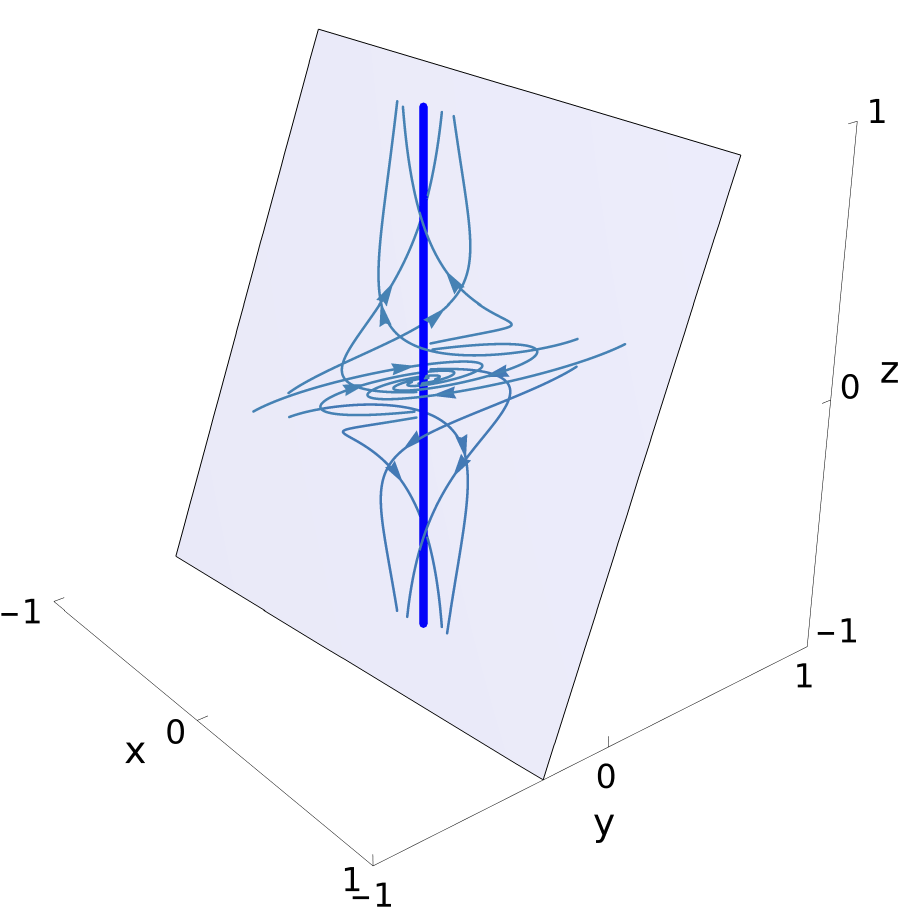}
    \caption{$k = 2;\, V_{11} = -0.5$}
    \label{fig:k2NSN}
  \end{subfigure}%

  \vspace{1ex}

   \begin{subfigure}[b]{0.33\textwidth}
    \includegraphics[width=\linewidth]{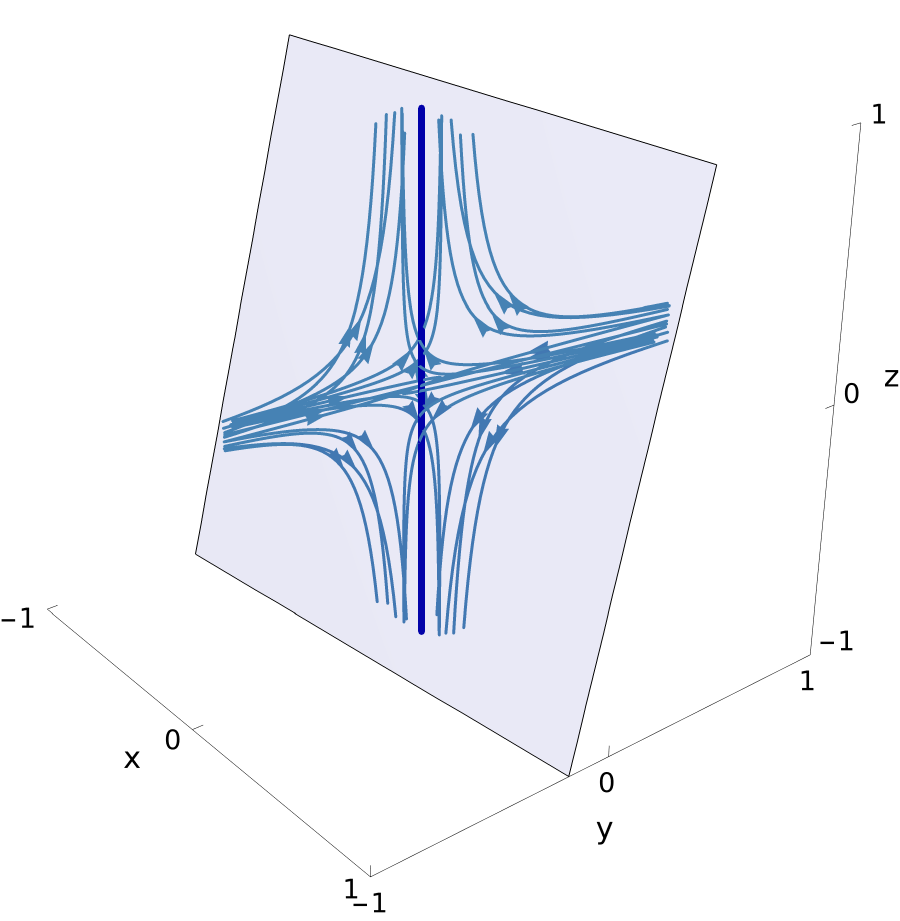}
    \caption{$k = 2;\, V_{11} = -0.43$}
    \label{fig:k2NRN}
  \end{subfigure}
  \hfill
  \begin{subfigure}[b]{0.33\textwidth}
    \includegraphics[width=\linewidth]{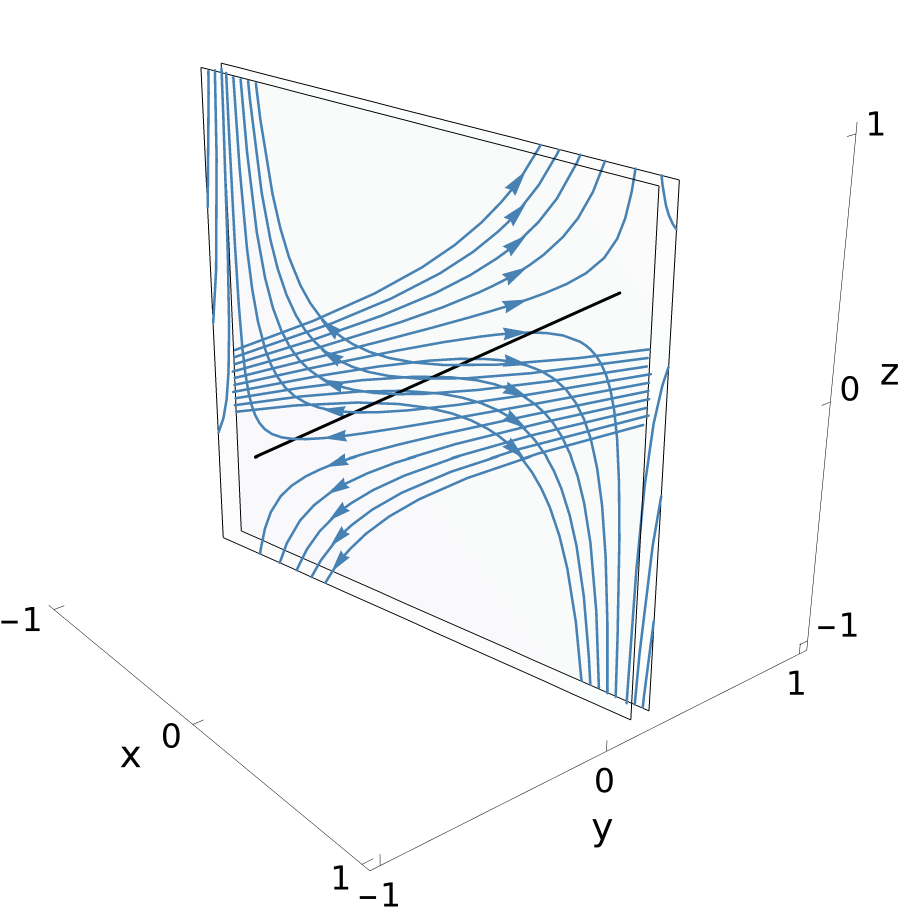}
    \caption{$k = 2;\, V_{11} = -0.4213$}
    \label{fig:k2Deg}
  \end{subfigure}%
  \hfill
  \begin{subfigure}[b]{0.33\textwidth}
    \includegraphics[width=\linewidth]{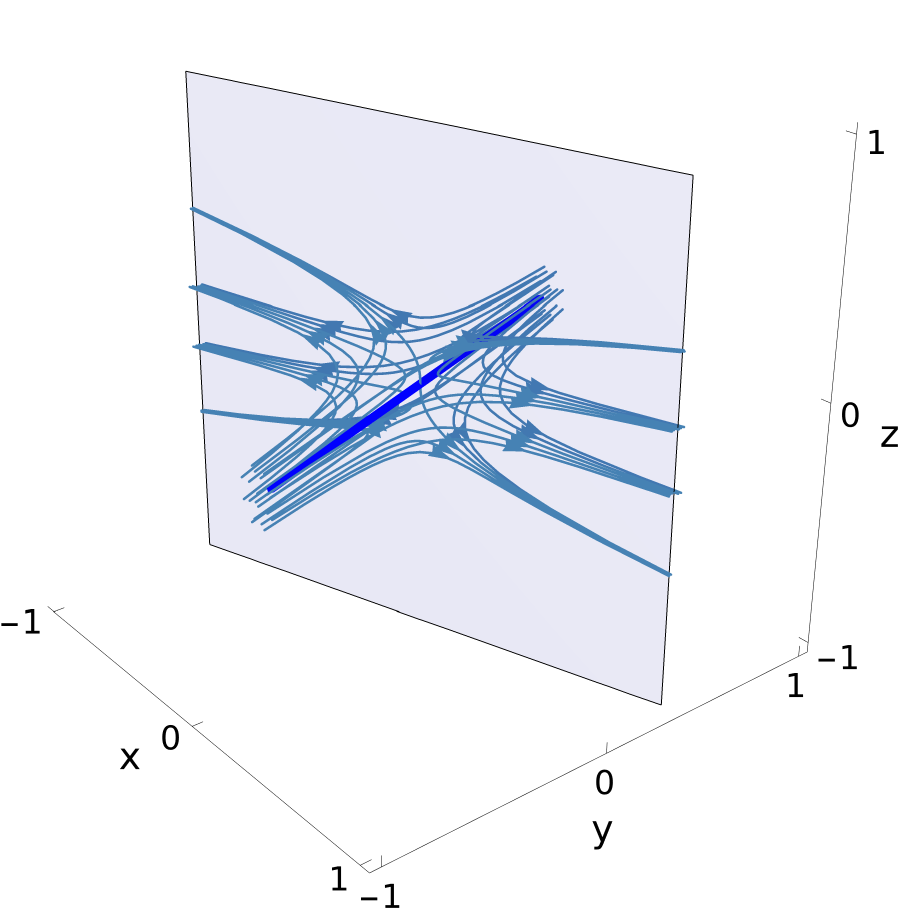}
    \caption{$k = 2;\, V_{11} = 1.0$}
    \label{fig:k2PRN}
  \end{subfigure}

  \caption{Panel~(a) shows the magnetic-field structure for $b=0.5$, $q=0.3$, $j_{\parallel}=0.1$, and $j_{\perp}=0.1$, corresponding to an improper radial null. Panels~(b)--(e) show velocity-field configurations for branch $k=1$, and panels~(f)--(i) those for branch $k=2$, along the $V_{13}=0$ slice. Panel~(c) shows the field-aligned solution, $\mathcal{V}=\mathcal{B}$. }
  \label{fig:ex1_flows}
\end{figure*}

Figure~\ref{fig:ex1_flows} shows representative velocity-field configurations, with $V_{11}$ selected to sample the distinct topological regimes identified above. For branch $k=1$, the antisymmetry about $V_{11}=0$ allows us to restrict the analysis to $V_{11}>0$. Panels~(b)--(e) therefore show, in sequence, a positive radial null at $V_{11}=0.5$, the field-aligned solution at $V_{11}=1.0$, the degenerate configuration at the bifurcation point $V_{11}=1.5898$, and a negative radial null at $V_{11}=2.0$. The field-aligned configuration is a special case of the general solution, occurring when $\mathcal{V}=\mathcal{B}$. Panels~(f)--(i) show the corresponding evolution along branch $k=2$. The sequence comprises a negative spiral null at $V_{11}=-0.5$, a negative radial null at $V_{11}=-0.43$ within the narrow interval between the spiral-to-radial transition and the bifurcation, the degenerate configuration at $V_{11}=-0.4213$, and a positive radial null at $V_{11}=1.0$. In each panel, the shaded surface represents the fan plane and the thick line shows the spine. In the degenerate cases, the black line denotes the line of nulls associated with the vanishing eigenvalue, so that the null is no longer isolated. In Figure~\ref{fig:k2Deg}, this line lies only $2.3^{\circ}$ from the invariant plane, which is the plane spanned by the eigenvectors corresponding to the two non-zero eigenvalues, within which the flow
is confined. The family of parallel invariant planes is therefore not visually distinguishable, and only two of the representative planes are shown.

The mechanism of this transition is instructive. Approaching the bifurcation point on branch $k=2$ from below, the eigenvalue associated with the minor fan axis tends to zero, where major and minor denote the eigenvectors associated with the larger and smaller eigenvalues in magnitude. As the inflow along this direction diminishes, the field lines in the fan plane increasingly converge toward the major axis. At the bifurcation, the minor eigenvalue vanishes identically and the isolated null is replaced by a line of nulls directed along the corresponding eigenvector. Since $\mathcal{V}$ annihilates this direction, the velocity field is invariant under translation along it, and the flow reduces to a two-dimensional hyperbolic configuration replicated in the family of parallel invariant planes spanned by the spine and the major fan axis (Figure~\ref{fig:k2Deg}). Beyond the bifurcation the minor eigenvalue changes sign and the null reorganizes into a three-dimensional structure of opposite polarity, where the new spine is directed along the former major fan axis, while the new fan plane is spanned by the former spine and minor fan directions. An analogous transition occurs on branch $k=1$. In both cases the reversal of polarity proceeds through an intermediate degenerate configuration in which one eigenvalue of the velocity Jacobian vanishes and the null is no longer isolated.

As a verification of our general solution, we consider the spiral magnetic-field configuration with $b=-0.5$, $q=-0.5$, $j_{\parallel}=2$, and $j_{\perp}=0.25$, for which $|j_{\parallel}|>j_c=\sqrt{(b-1)^2+q^2}$. In this case, our solutions recover the field-aligned (FA) and spiral field-crossing (SFC) flows identified by \citet{titov2000}. Specifically, the FA solutions occur at $V_{11}=\pm1$, while the SFC solutions are obtained at $V_{11}=\pm5\sqrt{2}/8$. In both cases, the resistivity is found to be constant, despite not being imposed a priori, demonstrating that the constant-resistivity analytical solutions of \citet{titov2000} are recovered as special cases of our more general formulation.

\subsection{Pressure Distribution}
\label{sec:pressure}

\begin{figure}[t!]
    \centering
    \includegraphics[width=\linewidth]{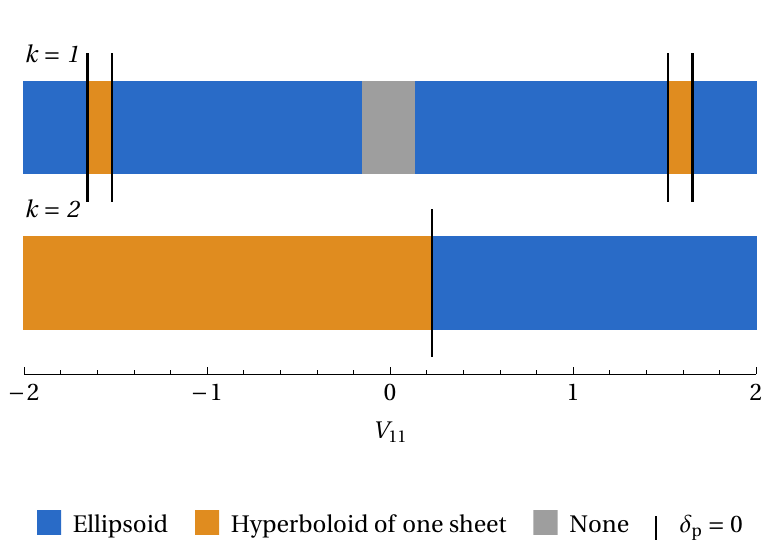}
    \caption{Classification of the pressure iso-surfaces along the slice $V_{13}=0$ for the solution branches $k=1$ (top) and $k=2$ (bottom). Colors denote the type of quadric surface according to the legend below. Vertical black lines mark the points at which $\delta_{p}$ vanishes. Gray regions denote intervals where the branch has no real solutions.}
    \label{fig:ex_pressure}
\end{figure}

For the magnetic-field configuration specified by Equation~\eqref{exB1}, substituting the corresponding velocity and magnetic-field Jacobians into Equation~\eqref{eq:pHess} yields the components of the pressure Hessian matrix as functions of $V_{11}$. These are then used in Equation~\eqref{eq:pexp} to obtain the pressure distribution. Here, $p_{0}$ denotes the pressure at the null point and must be positive. Since only the sign of $p_{0}$ affects the geometry of the isobaric surfaces, we set $p_{0}=1$ throughout the analysis. The pressure Hessian is symmetric $(\mathcal{P}_{ij}=\mathcal{P}_{ji})$, leaving six independent components. Along the slice $V_{13} = 0$ and fixed magnetic field parameters 
($b = 0.5,\,q = 0.3,\, j_{\parallel} = 0.1,$ and $j_{\perp} = 0.1$),
these reduce to
\begin{subequations}
\label{ex:PresHess}
\begin{align}
    \mathcal{P}_{11}^{(k)} &= - \tfrac{1}{50} -V_{11}^2 - V_{12}^{(k)}V_{21}^{(k)} , \\
    \mathcal{P}_{12}^{(k)} &=  -\tfrac{1}{20} - V_{12}^{(k)}(V_{11}+V_{22}^{(k)}), \\
    \mathcal{P}_{13}^{(k)} &= 0, \\
    \mathcal{P}_{22}^{(k)} &=  -V_{12}^{(k)}V_{21}^{(k)} - \left(V_{22}^{(k)}\right)^2, \\
    \mathcal{P}_{23}^{(k)} &= \tfrac{3}{20}, \\
    \mathcal{P}_{33}^{(k)} &= -\left(V_{33}^{(k)}\right)^2.
\end{align}
\end{subequations}

The pressure is a quadratic polynomial in the spatial coordinates. The isobaric surfaces are therefore second-order surfaces (quadrics), whose geometry is fully characterized by four invariants (see, e.g., \citealt{bartsch2014}):
\begin{subequations}
\begin{align}
\Delta^{(k)}_{p} &=
\begin{vmatrix}
\mathcal{P}_{11}^{(k)} & \mathcal{P}_{12}^{(k)} & \mathcal{P}_{13}^{(k)} & 0 \\
\mathcal{P}_{12}^{(k)} & \mathcal{P}_{22}^{(k)} & \mathcal{P}_{23}^{(k)} & 0 \\
\mathcal{P}_{13}^{(k)} & \mathcal{P}_{23}^{(k)} & \mathcal{P}_{33}^{(k)} & 0 \\
0 & 0 & 0 & p_{0}
\end{vmatrix}, \\
\delta^{(k)}_{p} &=
\begin{vmatrix}
\mathcal{P}_{11}^{(k)} & \mathcal{P}_{12}^{(k)} & \mathcal{P}_{13}^{(k)} \\
\mathcal{P}_{12}^{(k)} & \mathcal{P}_{22}^{(k)} & \mathcal{P}_{23}^{(k)} \\
\mathcal{P}_{13}^{(k)} & \mathcal{P}_{23}^{(k)} & \mathcal{P}_{33}^{(k)}
\end{vmatrix}, \\
t^{(k)}_{p} &=
\begin{vmatrix}
\mathcal{P}_{11}^{(k)} & \mathcal{P}_{12}^{(k)} \\
\mathcal{P}_{12}^{(k)} & \mathcal{P}_{22}^{(k)}
\end{vmatrix}
+
\begin{vmatrix}
\mathcal{P}_{22}^{(k)} & \mathcal{P}_{23}^{(k)} \\
\mathcal{P}_{23}^{(k)} & \mathcal{P}_{33}^{(k)}
\end{vmatrix}
+
\begin{vmatrix}
\mathcal{P}_{33}^{(k)} & \mathcal{P}_{13}^{(k)} \\
\mathcal{P}_{13}^{(k)} & \mathcal{P}_{11}^{(k)}
\end{vmatrix}, \\
s^{(k)}_{p} &= \mathcal{P}_{11}^{(k)} + \mathcal{P}_{22}^{(k)} + \mathcal{P}_{33}^{(k)}.
\end{align}
\end{subequations} 

The pressure contours may take the form of ellipsoids, hyperboloids of one sheet, or elliptical cylinders, depending on the values of the invariants. Ellipsoidal contours arise when $\Delta^{(k)}_{p} < 0$, $\delta^{(k)}_{p}s^{(k)}_{p} > 0$, and $t^{(k)}_{p} > 0$; the corresponding intervals of $V_{11}$ are shown in Figure~\ref{fig:ex_pressure}. One-sheet hyperboloids occur when $\Delta^{(k)}_{p} > 0$, with $\delta^{(k)}_{p}s^{(k)}_{p}$ and $t^{(k)}_{p}$ not both positive. The transition between the two regimes occurs through degenerate quadric surfaces corresponding to elliptical cylinders, characterized by $\delta^{(k)}_{p} = \Delta^{(k)}_{p} = 0$ with $t^{(k)}_{p} > 0$. On branch $k=1$ these occur at $V_{11} = \pm 1.5188$ and $V_{11} = \pm 1.6529$, bounding two narrow hyperboloidal windows, whereas on branch $k=2$ a single such transition occurs at $V_{11} = 0.2308$. In every case $t^{(k)}_{p}$ remains strictly positive across the transition, so the degenerate surface is an elliptical cylinder throughout.

As $\mathcal{P}^{(k)}_{33}\leq0$, the pressure Hessian is non-positive along the magnetic-field spine direction for every solution. Thus, $\mathcal{P}^{(k)}$ cannot be positive definite, and the pressure cannot have a local minimum at the null. The remaining possibilities are a local maximum, if $\mathcal{P}^{(k)}$ is negative definite with signature $(---)$, or a saddle, if it is indefinite with signature $(--+)$. These two cases correspond directly to the ellipsoidal and one-sheet hyperboloidal pressure iso-surfaces shown in Figure~\ref{fig:ex_pressure}: a negative-definite Hessian produces nested ellipsoids around a pressure maximum, whereas an indefinite Hessian produces hyperboloids centered on a saddle point. Because $(--+)$ is the only possible indefinite signature, at most one eigenvalue of $\mathcal{P}^{(k)}$ can be positive.

\subsection{Resistivity}

\begin{figure}
    \centering
    \includegraphics[width=\linewidth]{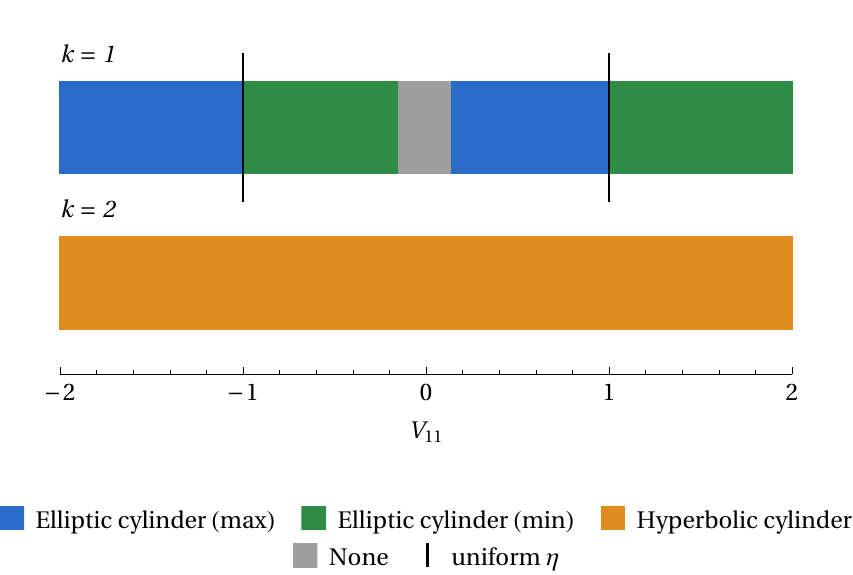}
    \caption{Geometry of the resistivity iso-surfaces along the $V_{13}=0$ slice for branches $k=1$ (top) and $k=2$ (bottom). Colors indicate the quadric surface type and whether the resistivity has a maximum or minimum at the null, as specified in the legend. Vertical black lines mark the points of spatially uniform resistivity. Gray regions denote intervals with no real solutions.}
    \label{fig:ex_resistivity}
\end{figure}
The general form of the expression for resistivity is given by Equation~\eqref{eq:rexp}. The explicit expressions for the resistivity obtained from Ohm's law, Equation~\eqref{eq:ohms}, are given in Equations~\eqref{eq:app_Res1} and \eqref{eq:app_Res2}, which provide the most general form considered here.
In this analysis, we restrict to $c_1=c_2=0$ in Equation~\eqref{eq:f_expansion}. For this choice, it follows that $\eta_{0} = -c_{3}$, the linear contribution vanishes, $\boldsymbol{\eta}_1=0$, and the remaining components of $\mathcal R$ are then determined by the magnetic field parameters and the two free parameters $V_{11}$ and $V_{13}$. Thus, for branch $k$, the resistivity can be written as
\begin{align}
    \eta^{(k)}
    ={}& \eta_0
    + \mathcal{R}^{(k)}_{11}x^2
    + 2\mathcal{R}^{(k)}_{12}xy
    + 2\mathcal{R}^{(k)}_{13}xz \nonumber\\
    &+ \mathcal{R}^{(k)}_{22}y^2
    + 2\mathcal{R}^{(k)}_{23}yz
    + \mathcal{R}^{(k)}_{33}z^2 .
\label{eq:Res}
\end{align}

For the magnetic-field configuration given by Equations~\eqref{exB1}-\eqref{exB1params}, and along the slice $V_{13}=0$, the components of the resistivity Hessian reduce to
\begin{subequations}
\label{eq:ResCoeff}
\begin{align}
    \mathcal{R}_{11}^{(k)} &= 2V_{11} - \tfrac{5}{2}V_{21}^{(k)} + V_{33}^{(k)} , \\
    \mathcal{R}_{12}^{(k)} &= \tfrac{1}{2}V_{21}^{(k)} - \tfrac{25}{2}V_{31}^{(k)} - V_{32}^{(k)} , \\
    \mathcal{R}_{13}^{(k)} &= 0 , \\
    \mathcal{R}_{22}^{(k)} &= -\tfrac{5}{2}V_{12}^{(k)} - V_{22}^{(k)} - \tfrac{1}{2}V_{33}^{(k)} , \\
    \mathcal{R}_{23}^{(k)} &= 0 , \\
    \mathcal{R}_{33}^{(k)} &= 0 .
\end{align}
\end{subequations}

Following the same procedure as for the pressure, we introduce the four invariants $\Delta_{\eta}^{(k)}$, $\delta_{\eta}^{(k)}$, $t_{\eta}^{(k)}$ and $s_{\eta}^{(k)}$ associated with the resistivity coefficients, which fully determine the geometry of the corresponding quadratic surfaces. Substituting Equations~\eqref{eq:ResCoeff} back into Equation~\eqref{eq:Res}, we find that the resistivity takes the simple form
\begin{equation}
    \eta^{(k)} = \mathcal{R}^{(k)}_{11}x^{2} + 2\mathcal{R}^{(k)}_{12}xy
                 + \mathcal{R}^{(k)}_{22}y^{2} + \eta_{0},
\end{equation}
so that $\eta^{(k)}$ depends only on the coordinates perpendicular to the magnetic spine, which is aligned with the $z$-axis. Writing $\mathbf{r}_{\perp}=(x,y)^{\mathrm{T}}$, the resistivity can therefore be expressed as
\begin{equation}
    \eta^{(k)} = \mathbf{r}_{\perp}^{\mathrm{T}}\,\mathcal{R}^{(k)}_{\perp}\,
                 \mathbf{r}_{\perp} + \eta_{0},
\end{equation}
where
\begin{equation}
    \mathcal{R}^{(k)}_{\perp}
    =
    \begin{pmatrix}
        \mathcal{R}^{(k)}_{11} & \mathcal{R}^{(k)}_{12} \\
        \mathcal{R}^{(k)}_{12} & \mathcal{R}^{(k)}_{22}
    \end{pmatrix}
\end{equation}
is the $2\times2$ submatrix of the resistivity Hessian $\mathcal{R}^{(k)}$ in the plane perpendicular to the magnetic spine.
The invariants reduce to $t_{\eta}^{(k)}=\det\mathcal{R}^{(k)}_{\perp}$ and $s_{\eta}^{(k)}=\operatorname{tr}\mathcal{R}^{(k)}_{\perp}$, and the resistivity is therefore independent of $z$ and is invariant along the magnetic spine. Consequently $\delta_{\eta}^{(k)}=\Delta_{\eta}^{(k)}=0$ identically, and the iso-surfaces are cylinders with their axes aligned with the magnetic spine.
The iso-surfaces are hence classified by the sign of the single remaining invariant $t_{\eta}^{(k)}=\mathcal{R}^{(k)}_{11}\mathcal{R}^{(k)}_{22} -\bigl(\mathcal{R}^{(k)}_{12}\bigr)^{2}$.  This is in contrast to the pressure, which retains a genuinely three-dimensional structure.

The resulting behavior is shown in Figure~\ref{fig:ex_resistivity}. On branch $k=1$, $t_{\eta}^{(1)}>0$ throughout, so the transverse form $\mathcal{R}^{(1)}_{\perp}$ is definite and the iso-surfaces are elliptic cylinders. The sign of $\mathcal{R}^{(1)}_{11}$ determines whether $\eta$ has a maximum or minimum at the null. At the field-aligned solutions, $V_{11}=\pm1$, all three coefficients vanish and the resistivity is uniform. For $0<V_{11}<1$, $\mathcal{R}^{(1)}_{11}<0$ and $\eta$ has a maximum at the null, whereas for $V_{11}>1$, $\mathcal{R}^{(1)}_{11}>0$ and $\eta$ has a minimum. Because each $\mathcal{R}^{(k)}_{ij}$ is linear in the components of $\mathcal{V}^{(k)}$, the coefficients are odd under $V_{11}\rightarrow -V_{11}$ on this branch, while $t_{\eta}^{(1)}$ is even. Thus, the surface classification is symmetric about $V_{11}=0$, with the maximum and minimum regimes exchanged. Branch $k=2$ is qualitatively different. Here, $t_{\eta}^{(2)}<0$ throughout, and the transverse form is therefore indefinite, giving hyperbolic-cylinder iso-surfaces and a saddle in $\eta$ at the null. This branch admits no field-aligned solution and hence no state of uniform resistivity.

The requirement $\eta>0$ imposes a further restriction on the physical domain. Since $\eta$ is independent of $z$, no additional constraint arises along the spine, unlike the pressure, for which $\mathcal{P}^{(k)}_{33}=-(V^{(k)}_{33})^2\leq0$ forces $p$ to decrease along the spine. In the transverse plane ($z = 0$), $\eta$ becomes negative at a finite distance from the null whenever $\mathcal{R}^{(k)}_{\perp}$ is not positive definite. For the hyperbolic case, the quadratic contribution vanishes along two straight lines through the null and changes sign across them. In the sectors where it is positive, $\eta$ increases away from the null, whereas in the sectors where it is negative, $\eta$ decreases and eventually becomes negative, restricting the physically admissible region to a bounded neighborhood of the null in those directions. In the elliptic-maximum case, $\eta$ decreases in every transverse direction and vanishes on an ellipse surrounding the null, so the solution is admissible only within this ellipse. Only the elliptic-minimum case, for which $\eta\geq\eta_{0}$ everywhere within the local model, is free of this restriction.

\section{Discussion}
\label{sec:discussion}

For our analysis, we have assumed that the magnetic null and the flow null are colocated. With this, we were able to compute the components of the velocity field in the vicinity of the magnetic null point (Figure~\ref{fig:ex1_flows}). In the following, we discuss the implications of the different solution spaces with respect to reconnection versus non-reconnection scenarios (Section~\ref{sect:recon}), reconnection flows in the solar (or stellar) corona (Section~\ref{sect:sol}), flows in the vicinity of the apex of astrospheres (Section~\ref{sect:astro}), and we also discuss our assumption of nonconstant resistivity (Section~\ref{sect:res}).

\subsection{Reconnection versus non-reconnection solutions}
\label{sect:recon}

Magnetic reconnection is ubiquitous in cosmic plasmas and is often connected with magnetic null points. For magnetic reconnection to occur in 2D, the existence of a magnetic null point (X-point) and associated separatrices is prerequisite \citep{2007mare.book.....P}. Magnetic reconnection not only serves to convert magnetic energy into particle energy, i.e. thermal energy and kinetic energy, but its signature is the new connection of magnetic flux and the possibility that the plasma flow is able to cross magnetic boundaries (separatrices), where the electrical resistivity is non-zero (at least locally around the null). Having no resistivity, like in ideal MHD, plasma flows cannot cross magnetic separatrices, hence no magnetic reconnection can take place.
Moreover, in the stationary case, the electric field in $z$-direction is constant. As the convective part of the total electric field vanishes at the magnetic null, the electric field is directly related to the resistivity and the current density at the null point. Therefore, the $z$-component of the electric field provides a measure for the magnetic reconnection, meaning that it can be regarded as reconnection rate.


In 3D, the reconnection scenario is more complex. In configurations without
null points, reconnection takes place in a localized nonideal region in which
the electric field has a component parallel to the magnetic field,
$E_\parallel \neq 0$. In this framework of \emph{general magnetic reconnection}
\citep{1988JGR....93.5559H, schindler1988}, the reconnection rate is given by the line integral $\int E_\parallel\, \mathrm{d}s$ along the magnetic field
lines threading the nonideal region. The flux-weighted sum of these line
integrals, $\int \mathbf{E}\cdot\mathbf{B}\, \mathrm{d}V$, determines the rate of
change of the total magnetic helicity. For reconnection to occur,
regions with sheared magnetic fields and localized strong currents in small domains must exist \citep{2003JGRA..108.1285P, 2006SoPh..235..191N, 2006ASTRA...2...63N}. Another way of determining the magnetic reconnection rate is to compute the path integrals along a specific (set of) field line(s) or the maximum of all possible electric field line integrals, passing through this resistive domain and ending at its boundaries \citep[see, e.g.,][]{2015PhPl...22d2117W}.

For a region with a null point in 3D, the before mentioned integrals can also be used
to calculate the magnetic reconnection rate because the resistive domain is bounded by a prescribed resistivity \citep{pontin2004, pontin2005}. The situation in our analysis is different, because the resistivity is calculated self-consistently with a local solution of all other physical variables. This leads to an unbounded magnetic field along with an unbounded resistive region. Consequently, we cannot specify the exact measure of the reconnection rate but only calculate the line integral for an artificially specified field line. Instead of using an unspecific (or not well defined) magnetic reconnection rate, the classical criterion of crossing the fan plane (= separatrix) can be used for identifying whether reconnection takes place or not. For the $V_{13} = 0$ slice studied in Section~\ref{sec:results}, the solutions fulfil the aforementioned criterion, except for the FA and SFC cases. The latter can be identified with a non-reconnective solution, for example, for the apex region of an astrosphere where the hydropause is identical to the magnetopause. 

Unlike classical reconnection scenarios, our nonideal region is not confined to a thin current sheet embedded in an ideal domain. The usual argument for approximate helicity conservation, namely that helicity dissipation is small compared with energy dissipation when current is concentrated in thin layers \citep{taylor1974, berger1984}, does not apply here since the current density in our solutions is uniform. For our solutions, the resistive contribution to the rate of change of magnetic helicity is given by $-2\int_{D} \eta\, \mathbf{j}\cdot\mathbf{B}\, \mathrm{d}V$, where $D$ is a region centered on the null. Given the ansatz $c_1 = c_2 = 0$ in Equation~(\ref{eq:f_expansion}), the resistivity contains only constant and quadratic terms, so that, with $\mathbf{B}$ linear in position and $\mathbf{j}$ uniform, the integrand is a sum of linear and cubic terms that are odd under $\mathbf{r} \rightarrow -\mathbf{r}$, and its integral over $D$ vanishes. In contrast, the Ohmic dissipation,
$\int_{D} \eta\, j^{2}\, \mathrm{d}V$, has an even integrand and is nonzero, so that magnetic energy is dissipated while magnetic helicity is not.

\subsection{Stellar (solar) coronae}
\label{sect:sol}

Negative nulls are required to describe the fan-spine magnetic topology of coronal jets
\citep{2016ApJ...827....4W, 2016ApJ...820...77W}. The fan--spine topology invoked in models of coronal jets specifies the magnetic configuration but not the accompanying flow. Our analysis shows that this distinction is quantifiable, i.e., for a given magnetic null, the stationary flows form a two-parameter family rather than a unique solution, with different members exhibiting qualitatively distinct topological skeletons. For the positive radial null considered in Section~\ref{sec:results}, the admissible velocity nulls include positive radial, negative radial, and spiral configurations, depending on the branch and $V_{11}$. Thus, identifying a fan--spine magnetic topology observationally does not by itself determine whether the associated stagnation flow is inflowing or outflowing, or whether it is rotational.

This distinction matters for jets in two respects. First, the sign of $\det\mathcal{V}$ determines the direction of the stagnation flow, specifying whether the plasma converges toward the null along the spine and diverges within the fan, or vice versa, and hence which direction along the open spine the outflow can be directed. The polarity of the velocity null is therefore a physically distinct property from that of the magnetic null. Second, the spiral branches possess non-zero vorticity that is not parallel to the current density, indicating rotation about the spine independently of magnetic twist. Since rotational motions are widely reported in coronal jets \citep[e.g.,][]{shen2011,chen2012,panesar2025, huang2026}
our solutions show that such motions can be sustained locally at a null even when the magnetic field itself is of radial rather than spiral type.

Finally, the results of Section~\ref{sect:recon} bear directly on the
interpretation of jets as reconnection events. Because the fan planes of the magnetic and velocity fields coincide only at the specific FA and SFC solutions, the generic stationary flow crosses the magnetic fan surface while remaining confined along the spine. The fan separatrix that organizes the jet geometry is therefore not, in general, a barrier to the plasma flow. Cross-separatrix transport follows directly from the solutions and does not require an additional physical assumption.

\subsection{Apex region of astrospheres}
\label{sect:astro}

Not only magnetic structures in stellar coronae or magnetospheres can have null points. 
Astrospheres can also be regarded as stages where null points of physical vector fields, such as magnetic field and flow field, occur. In general, it is assumed that astrospheres are marked by a stagnation point, which acts as navel for the so called astropause. This can also be regarded as a definition of the astropause. Accordingly, the magnetopause can be considered as a magnetic separatrix having the magnetic null point as a navel. However, as in the case of the heliopause \citep[e.g.,][]{2023FrASS..1060618L}, such kind of definitions has some ambiguity. In particular, the question that emerges is what is the role of the magnetopause concerning the global structure of the astrosphere.

The null points can be regarded as navels as they mark the intersection between a 1D manifold (spine curve: stagnation curve or magnetic spine curve) with the corresponding 2D manifold (fan surfaces: astropause or magnetopause). For a linear system, the curves are straight lines and the surfaces are planes. So one can regard the fan planes of the corresponding vector fields as tangential plane of the corresponding pauses. In the case of MHD, if both null points are colocated, the fans do not necessarily have to coalign but can be tilted so that both diverging fans locally define a finite-width boundary layer. Such cases can be interpreted as magnetic reconnection. 
But even if the fans locally coalign, they must not do so globally. In this case, 
magnetic reconnection or annihilation processes can take place at the flanks of the astrosphere.
An interesting case for future studies is the question how an astrosphere might look like if 
the null points are dislocated locally but are placed on the same hull, i.e., a common surface (pause). This would require a specific configuration of the nonideal term.

To describe astrospheres as boundary layers between the inflowing interstellar wind and an outflowing stellar wind, the stagnation point must have negative eigenvalues concerning the 1D submanifold meaning that it is a positive null. This guarantees that both flows converge towards and encounter at the stagnation point. In our analysis, we have assumed that the stagnation point and the null point co-locate. Under this constraint, we identify the ranges of $V_{11}$ along the slice $V_{13}=0$ for which the velocity null is of positive type (Figure~\ref{fig:ex1Bfc}). Within these ranges, the pressure at the null can correspond either to a local maximum or to a saddle point, as discussed in Section~\ref{sec:pressure}.
Hence, our results allow us to describe a large set of astrosphere configurations at least locally.

\subsection{The nonconstant resistivity}
\label{sect:res}

In our treatment, the resistivity is not necessarily constant. 
As previous studies have shown, a constant resistivity implies that the variability of the nonideal term depends purely on the variation of the current density. This means that the reconnection rate or dissipation rate depends explicitly on the current density only. While from the physical point of view  this is what one would expect, it is also a paradox, because a microscopic value namely a constant resistivity implies that the typical interaction frequency or collision frequency is not spatially varying, while the macroscopic variables like magnetic field and velocity fields provide a gradient close to their null points. The interaction frequency or collision frequency is functionally related to the resistivity \citep[see, e.g.,][]{1977RvGSP..15..113P}. A complication arising from a constant resistivity is that under certain conditions, for example for sub-alfv\'{e}nic flows, no reconnection can occur, which is known as the anti-reconnection theorem \citep{titov2000, 2007mare.book.....P}\footnote{Anti-reconnection theorems state that either no separatrix can be crossed by the bulk plasma flow, or only specific subsets of the manifolds determining the separatrices \citep{craig1995, nickeler2012}.}.

To facilitate reconnection or dissipation, often the concept of anomalous resistivity is employed, which means that the resistivity is an explicit function of the current density, $\mathbf{j}$. This implies that within a domain, where the current is highly concentrated, the reconnection rate is enhanced and can trigger current-driven instabilities. However, taking into account smaller macroscopic scales, for example within the center of a macroscopic current, the resistivity is also constant and can locally not contribute to the reconnection rate. 

Dropping the constant resistivity constraint implies that we calculate the resistivity, the magnetic field, the flow field, and the pressure self-consistently and simultaneously, allowing to consider the mutual nonlinear feedback of every variable. Two consequences of this are worth emphasizing.

First, the present formulation makes explicit why an anomalous resistivity prescription cannot generate spatial structure in the present linear regime. Because the magnetic field is linear in the spatial coordinates, its current density is spatially uniform. Thus, any prescription of the form $\eta=\eta(|\mathbf{j}|)$ yields a constant resistivity near the null. By contrast, the resistivity obtained here varies quadratically with position, even though the current density is uniform. Its spatial structure is therefore determined by the velocity field rather than by current concentration. In particular, the quadratic coefficients $\mathcal{R}^{(k)}_{ij}$ vanish for the field-aligned solutions $\mathcal{V}=\pm\mathcal{B}$, for which $\eta=\eta_0$, so that the departure of the resistivity from uniformity directly reflects the departure of the flow from field alignment.

Second, the constant-resistivity FA and SFC solutions are precisely those for which $\mathcal{V}\mathcal{B}=\mathcal{B}\mathcal{V}$. The two matrices then share a common eigenbasis, so their spines and fan planes coincide and the plasma remains confined by the magnetic separatrices. For the generic
solutions obtained with nonconstant resistivity, the commutator is non-zero and the magnetic and velocity fan planes are misaligned, allowing the flow to cross the magnetic separatrix. Thus, within the present
formulation, relaxing the constant-resistivity constraint admits generic separatrix-crossing solutions, while the constant-resistivity FA and SFC solutions remain non-reconnective.

We would like to emphasize that a spatially varying resistivity is not necessarily unphysical, provided that it is interpreted as an effective representation of the non-ideal term in Equation~\eqref{eq:ohm} rather than necessarily as a collisional transport coefficient. Investigations of plasma dynamics around magnetic nulls reveal that nonideal effects can arise from processes other than classical resistivity, including electron pressure effects and collisionless particle dynamics, and can be localized around the null and structured by the local magnetic-field geometry instead of by the current density alone \citep{tsiklauri2008,he2008,hesse2016,shang2017}. In particular, \citet{shang2017} demonstrated that collisionless particle dynamics near a magnetic null can produce an effective resistive description within a finite region around the null, with its strength depending on the local field geometry and not on the current density alone. These results show that spatially structured effective non-idealness can arise from underlying microphysical processes, and therefore that a nonuniform effective resistivity is not, in itself, inconsistent with the physics of magnetic nulls. However, we do not claim that the particular spatial profile obtained in our solutions follows from any specific microscopic transport mechanism.

Importantly, our solutions do not prescribe the magnitude of the resistivity. With $c_1 = c_2 = 0$ in Equation~\eqref{eq:f_expansion}, the resistivity at the null is $\eta_0 = -c_3$, where $c_3$ is an arbitrary constant. The stationary equations therefore constrain the spatial variation of the effective resistivity relative to its value at the null, but do not determine that value itself. Since both the velocity and the magnetic field vanish at the null, Ohm's law, Equation~\eqref{eq:ohm}, gives $\mathbf{E}(\mathbf{0}) = \eta_0 \mathbf{j}$, so that the microphysics remains responsible for setting the magnitude of the nonideal electric field and hence the overall level of dissipation. The equations hence identify mutually compatible stationary configurations, while the underlying microphysical transport processes, together with additional physical constraints not included in the present local MHD model, would determine which member of this family can be realized. A specific microphysical closure for 3D magnetic nulls is beyond the scope of the present work, and we therefore do not claim to determine which member of the family is selected in a particular physical plasma.

\section{Conclusions and Outlook}
\label{sec:conclusions}

We have developed the solution space of the local 3D resistive, incompressible MHD system for the case where magnetic null and velocity null coincide. 
Our analysis enables us to identify the linkage between the vector field lines amongst each other as well as their \lq knotting\rq \ around the null point itself. With this, our solutions reflect the topological skeleton of the entire 3D resistive MHD system, which is represented by the eigenvalues of all involved fields. Our calculations have no predefined symmetry, so the configurations considered are fully 3D and generic. The resistivity is self-consistently calculated together with the topological properties of the magnetic field, the flow field, and the pressure distribution. Consequently, the resistivity can be considered as a microscopic physical variable, which is not prescribed from external parameters but from the solution itself, reflecting the nonlinear feedback between microscopic transport parameters and macroscopic variables. 

The resulting solution space shows that a given magnetic null does not uniquely determine the associated stationary flow. Instead, a two-parameter family of solutions is obtained, which exhibits distinct flow topologies, pressure structures, and resistivity distributions, demonstrating the coupled nature of these quantities in the full MHD system.

For a fixed positive improper radial magnetic null, we find that the admissible flows include positive radial, negative radial, and spiral topologies. The flow polarity reverses at bifurcation points, where the velocity field becomes effectively $2.5$D. The spiral solutions introduce rotational flow that is absent from the magnetic topology itself. More importantly, the generic solutions do not require the velocity and magnetic separatrices to coincide, allowing plasma to cross the magnetic separatrix. For specific choices of the free parameters, we recover the FA and SFC solutions of \citet{titov2000} with uniform resistivity.

These results demonstrate that the local topology of a magnetic null alone does not determine the surrounding plasma dynamics. The self-consistent treatment of resistivity reveals a broader class of stationary 3D flows, including distinct inflow, outflow, and rotational configurations and generic separatrix-crossing solutions. The bifurcations are transitions that also mark the distinction between the scenarios \lq accretion\rq \ (negative flow nulls) and \lq decrease \rq \ (positive flow nulls). The latter are particularly relevant for the definition of astrospheric flows.

A natural next step is a perturbative stability analysis of the family of stationary solutions, to determine which of its members are stable. This analysis could be complemented by numerical simulations initialized with a member of the solution family, including the corresponding resistivity distribution, to examine whether the configuration persists. Simulations with a fixed magnetic null and different imposed flows would further test whether distinct flow topologies can be sustained around the same null. 
The solutions could also be tested observationally. For example, in-situ measurements in the terrestrial magnetosphere determine the magnetic-field Jacobian at 3D nulls \citep[e.g.,][]{guo2022}, and simultaneous plasma measurements provide the velocity Jacobian, allowing the topologies of the velocity and magnetic nulls to be compared directly. In the laboratory, recently developed techniques, such as coil assemblies that produce controllable 3D nulls with sustained strong fields \citep[e.g.,][]{chesny2020,chesny2022} offer further possibilities to test the existence of our solutions, provided that the magnetic and flow fields near the null can be measured simultaneously.

\begin{acknowledgments}
We would like to thank the reviewer for valuable comments.
This research made use of the NASA Astrophysics Data System (ADS).
K.S.S., D.H.N. and M.K. acknowledge financial support from the Czech Science Foundation 
(GA \v{C}R, grant number 25-17532S). The Astronomical Institute of the Czech Academy of Sciences is supported by the project RVO:67985815. This project is co-funded by the European Union (Project 101183150 - OCEANS). 
\end{acknowledgments}

\bibliography{ms}{}
\bibliographystyle{aasjournalv7}

\begin{appendix}

\section{Derivation of the stationary solutions}
\label{app:derivation}
This appendix provides the algebraic details omitted from Section~\ref{sec:methods}. The solution procedure follows the steps outlined in the main text. The electrostatic potential is obtained from Ohm's law, the resistivity is subsequently determined, and the resulting consistency conditions are used together with the stationary momentum equation to determine the admissible velocity Jacobian elements.

\subsection{The Electrostatic Potential}
The electrostatic potential is constructed from the system of Ohm's law equations, Equation~\eqref{eq:ohms}. Integrating Equation~\eqref{eq:ohms_b} with respect to $y$ yields
\begin{equation}
    \phi(x,y,z) = F(x,y,z) + g(x,z),
\end{equation}
where the polynomial contribution is
\begin{align}
	F(x,y,z) & \nonumber = x^{2}y\left[ V_{31} \right] + xy^{2}\left[ \frac{1}{2}\frac{(q - j_{\parallel})}{2}V_{31} + \frac{1}{2}V_{32} - \frac{j_{\perp}}{2} V_{11} \right]
	+ y^{3}\left[ \frac{1}{3}\frac{(q - j_{\parallel})}{2}V_{32} - \frac{j_{\perp}}{3} V_{12} \right] + xyz\left[ (b+1)V_{11} + V_{33} \right] \\ & + y^{2}z\left[ \frac{(b+1)}{2}V_{12} 
	+ \frac{1}{2}\frac{(q - j_{\parallel})}{2}V_{33} - \frac{j_{\perp}}{2}V_{13} \right] + yz^{2} \left[ (b+1)V_{13} \right],
\end{align}
and $g(x,z)$ is an arbitrary function of $x$ and $z$. The function $g(x,z)$ is obtained by eliminating the resistivity from Equations~\eqref{eq:ohms_a} and \eqref{eq:ohms_c}. Multiplying Equation~\eqref{eq:ohms_a} by $j_{\parallel}$, Equation~\eqref{eq:ohms_c} by $j_{\perp}$, and subtracting the resulting equations leads to
\begin{align}
	& \nonumber j_{\parallel}\frac{\partial g(x,z)}{\partial x} - j_{\perp}\frac{\partial g(x,z)}{\partial z} = xy\Bigl[ j_{\perp}(V_{11} + V_{22} + V_{33}) 
	-j_{\perp}\frac{(q+j_{\parallel})}{2}V_{12} + j_{\perp}\frac{(q+j_{\parallel})}{2}V_{21} - j_{\parallel}(b+2)V_{31} - j_{\parallel}\frac{(q+j_{\parallel})}{2}V_{32} \Bigr] \\ \nonumber & + y^{2}\Bigl[ -\frac{j_{\perp}}{2}(b-1)V_{12} + j_{\perp}\frac{(q+j_{\parallel})}{2}V_{22}  - \frac{j_{\parallel}}{2}(2b+1)V_{32} - \frac{j_{\parallel}}{2}\frac{(q-j_{\parallel})}{2}V_{31} +\frac{j_{\parallel}j_{\perp}}{2}V_{11} +\frac{j_{\perp}^{2}}{2}V_{13}  +\frac{j_{\perp}}{2}\frac{(q-j_{\parallel})}{2}V_{33}  \Bigr] \\ \nonumber & + yz \Bigl[ j_{\perp}(b+2)V_{13} +j_{\perp}\frac{(q+j_{\parallel})}{2}V_{23} - j_{\parallel}(b+1)(V_{11} + V_{22} + V_{33}) \Bigr] + x^{2}\Bigl[ -j_{\perp}\frac{(q+j_{\parallel})}{2}V_{11} + j_{\perp}V_{21} - j_{\parallel}\frac{(q+j_{\parallel})}{2}V_{31} \Bigr] \\  & + xz \Bigl[ -j_{\perp}\frac{(q+j_{\parallel})}{2}V_{13} -j_{\parallel}(b+1)V_{21}  + j_{\perp}V_{23}  - j_{\parallel}\frac{(q+j_{\parallel})}{2}V_{33} \Bigr] + z^{2}\Bigl[ -j_{\parallel}(b+1)V_{23} \Bigr].
\end{align}
Since $g(x,z)$ is independent of $y$, the right-hand side of Equation~\eqref{eq:g_PDE} must likewise be independent of $y$. Consequently, the coefficients of the $xy$, $y^2$, and $yz$ terms must vanish identically, giving the three linear constraints in velocity Jacobian elements
\begin{align}
  -j_{\perp}(V_{11} + V_{22} + V_{33}) +j_{\perp}\frac{(q+j_{\parallel})}{2}V_{12} - j_{\perp}\frac{(q+j_{\parallel})}{2}V_{21}  + j_{\parallel}(b+2)V_{31} + j_{\parallel}\frac{(q+j_{\parallel})}{2}V_{32} &= 0, \label{eq:vcons_phi_1} \\  
	\frac{j_{\perp}}{2}(b-1)V_{12} - j_{\perp}\frac{(q+j_{\parallel})}{2}V_{22} + \frac{j_{\parallel}}{2}(2b+1)V_{32}-\frac{j_{\parallel}j_{\perp}}{2}V_{11}  + \frac{j_{\parallel}}{2}\frac{(q-j_{\parallel})}{2}V_{31}  -\frac{j_{\perp}^{2}}{2}V_{13} -\frac{j_{\perp}}{2}\frac{(q-j_{\parallel})}{2}V_{33} &=0, \label{eq:vcons_phi_2} \\  
	-j_{\perp}(b+2)V_{13} + j_{\parallel}(b+1)(V_{11} + V_{22} + V_{33}) -j_{\perp}\frac{(q+j_{\parallel})}{2}V_{23} & = 0. \label{eq:vcons_phi_3}
\end{align}
Under these conditions, the source term  $S(x,z)$ reduces to
\begin{align}
    S(x,z) & \nonumber=  x^{2}\Bigl[ -j_{\perp}\frac{(q+j_{\parallel})}{2}V_{11} + j_{\perp}V_{21} - j_{\parallel}\frac{(q+j_{\parallel})}{2}V_{31} \Bigr] + xz \Bigl[ -j_{\perp}\frac{(q+j_{\parallel})}{2}V_{13} -j_{\parallel}(b+1)V_{21}  + j_{\perp}V_{23}  - j_{\parallel}\frac{(q+j_{\parallel})}{2}V_{33} \Bigr] \\ & + z^{2}\Bigl[ -j_{\parallel}(b+1)V_{23} \Bigr].
\end{align}
The resulting first-order PDE is readily solved by the method of characteristics. The polynomial particular solution $G(x,z)$ introduced in Equation~\eqref{eq:g_sol} is
\begin{align}
	G(x,z) & \nonumber = x^{3}\Bigl[ -\frac{j_{\perp}^{2}}{6j_{\parallel}^{2}} \frac{(q+j_{\parallel})}{2}V_{13} - \frac{j_{\perp}^{2}}{6j_{\parallel}^{2}}(1+2b)V_{23}   - \frac{1}{3}\frac{(q+j_{\parallel})}{2}V_{31} - \frac{j_{\perp}}{6j_{\parallel}}(b-1)V_{21} - \frac{j_{\perp}}{3j_{\parallel}}\frac{(q+j_{\parallel})}{2}V_{11} - \frac{j_{\perp}}{6j_{\parallel}}\frac{(q+j_{\parallel})}{2}V_{33} \Bigr] \\ & + x^{2}z \Bigl[ -\frac{1}{2}\frac{(q+j_{\parallel})}{2}V_{33}  - \frac{j_{\perp}}{2j_{\parallel}}(1+2b)V_{23} - \frac{j_{\perp}}{2j_{\parallel}}\frac{(q+j_{\parallel})}{2}V_{13} - \frac{1}{2}(b+1)V_{21} \Bigr] + xz^{2} \Bigl[- (b+1)V_{23} \Bigr],
\end{align}
Combining the polynomial contributions $F(x,y,z)$ and $G(x,z)$ with the Taylor expansion of the arbitrary function $f$, Equation~\eqref{eq:f_expansion}, the electrostatic potential assumes the form
\begin{align}
	\phi(x,y,z) & \nonumber = x^{2}y\left[ V_{31} \right] + xy^{2}\left[ \frac{1}{2}\frac{(q - j_{\parallel})}{2}V_{31} + \frac{1}{2}V_{32} - \frac{j_{\perp}}{2} V_{11} \right]
	+ y^{3}\left[ \frac{1}{3}\frac{(q - j_{\parallel})}{2}V_{32} - \frac{j_{\perp}}{3} V_{12} \right] + xyz\left[ (b+1)V_{11} + V_{33} \right] \\ \nonumber & + y^{2}z\left[ \frac{(b+1)}{2}V_{12} 
	+ \frac{1}{2}\frac{(q - j_{\parallel})}{2}V_{33} - \frac{j_{\perp}}{2}V_{13} \right] + yz^{2} \left[ (b+1)V_{13} \right] + xz^{2} \Bigl[  - (1+b)V_{23} +3c_{1}j_{\perp}j_{\parallel}^2 \Bigr] \\ \nonumber & + x^{3}\Bigl[ -\frac{j_{\perp}^{2}}{6j_{\parallel}^{2}} \frac{(q+j_{\parallel})}{2}V_{13} - \frac{j_{\perp}^{2}}{6j_{\parallel}^{2}}(1+2b)V_{23}  - \frac{1}{3}\frac{(q+j_{\parallel})}{2}V_{31} - \frac{j_{\perp}}{6j_{\parallel}}(b-1)V_{21} - \frac{j_{\perp}}{3j_{\parallel}}\frac{(q+j_{\parallel})}{2}V_{11}  - \frac{j_{\perp}}{6j_{\parallel}}\frac{(q+j_{\parallel})}{2}V_{33} \\ \nonumber & + c_{1}j_{\perp}^{3} \Bigr]  + x^{2}z \Bigl[ -\frac{1}{2}\frac{(q+j_{\parallel})}{2}V_{33}  - \frac{j_{\perp}}{2j_{\parallel}}(1+2b)V_{23} - \frac{j_{\perp}}{2j_{\parallel}}\frac{(q+j_{\parallel})}{2}V_{13} -\frac{1}{2}(b+1)V_{21} +3c_{1}j_{\perp}^{2}j_{\parallel}  \Bigr]   + z^{3}\Bigl[ c_{1}j_{\parallel}^{3} \Bigr]  \\ & + x^{2}\Bigl[ c_{2}j_{\perp}^2 \Bigr]  + xz\Bigl[ 2c_{2}j_{\parallel}j_{\perp} \Bigr] + z^{2}\Bigl[ c_{2}j_{\parallel}^{2} \Bigr]+ x\Bigl[ c_{3}j_{\perp} \Bigr] + z\Bigl[ c_{3}j_{\parallel} \Bigr] + c_{4}.
    \label{eq:app_EPot}
\end{align}

\subsection{Resistivity}

The resistivity is obtained by substituting the electrostatic potential into Ohm's law. Using the $x$-component, Equation~\eqref{eq:ohms_a}, and solving for $\eta$ gives
\begin{align}
	\eta(x,y,z) & \nonumber = x^{2}\Bigl[ \frac{j_{\perp}}{2j_{\parallel}^{2}}\frac{(q+j_{\parallel})}{2}V_{13} + \frac{j_{\perp}}{2j_{\parallel}^{2}}(1+2b)V_{23}+ \frac{1}{2j_{\parallel}}(b-1)V_{21}+ \frac{1}{j_{\parallel}}\frac{(q+j_{\parallel})}{2}V_{11}  + \frac{1}{2j_{\parallel}}\frac{(q+j_{\parallel})}{2}V_{33} -3 c_{1}j_{\perp}^{2} \Bigr] \\ \nonumber & + xy \Bigl[ 
	V_{21}- \frac{1}{j_{\perp}}(b+2)V_{31} - \frac{1}{j_{\perp}} \frac{(q+j_{\parallel})}{2}V_{32} \Bigr] + xz \Bigl[ \frac{1}{j_{\parallel}}(1+2b)V_{23}+ \frac{1}{j_{\parallel}}\frac{(q+j_{\parallel})}{2}V_{13} - 6c_{1}j_{\parallel}j_{\perp} \Bigr] \\ \nonumber & + y^{2} \Bigl[ 
	V_{22} - \frac{1}{2j_{\perp}}\frac{(q-j_{\parallel})}{2}V_{31}- \frac{1}{2j_{\perp}}(1+2b)V_{32} + \frac{1}{2}V_{11} \Bigr] + z^{2}\Bigl[ -3c_{1}j_{\parallel}^{2} \Bigr] + yz\Bigl[ V_{23} -\frac{(b+1)}{j_{\perp}}V_{11} + V_{22} + V_{33} \Bigr] \\ & + x\Bigl[ -2c_{2}j_{\perp}\Bigr] + z\Bigl[ -2c_{2}j_{\parallel} \Bigr] - c_{3}.
    \label{eq:app_Res1}
\end{align}
An independent expression follows from the $z$-component of Ohm's law, Equation~\eqref{eq:ohms_c},
\begin{align}
	\eta(x,y,z) & \nonumber = x^{2}\Bigl[ \frac{j_{\perp}}{2j_{\parallel}^{2}}\frac{(q+j_{\parallel})}{2}V_{13} + \frac{j_{\perp}}{2j_{\parallel}^{2}}(1+2b)V_{23}+ \frac{1}{2j_{\parallel}}(b-1)V_{21}+ \frac{1}{j_{\parallel}}\frac{(q+j_{\parallel})}{2}V_{11}  + \frac{1}{2j_{\parallel}}\frac{(q+j_{\parallel})}{2}V_{33} -3 c_{1}j_{\perp}^{2} \Bigr] \\ \nonumber & + xy \Bigl[ 
	\frac{1}{j_{\parallel}} \frac{(q+j_{\parallel})}{2}V_{12} - \frac{1}{j_{\parallel}}\frac{(q-j_{\parallel})}{2}V_{21} -\frac{1}{j_{\parallel}}(V_{11} + V_{22} + V_{33})\Bigr] + xz \Bigl[ \frac{1}{j_{\parallel}}(1+2b)V_{23}+ \frac{1}{j_{\parallel}}\frac{(q+j_{\parallel})}{2}V_{13} - 6c_{1}j_{\parallel}j_{\perp} \Bigr]\\ \nonumber & + y^{2} \Bigl[ \frac{1}{2j_{\parallel}}(b-1)V_{12} - \frac{1}{j_{\parallel}}\frac{(q-j_{\parallel})}{2}V_{22}- \frac{1}{2j_{\parallel}}\frac{(q-j_{\parallel})}{2}V_{33}+ \frac{1}{2}\frac{j_{\perp}}{j_{\parallel}}V_{13} \Bigr]  + yz\Bigl[ -\frac{1}{j_{\parallel}}(b+2)V_{13} - \frac{1}{j_{\parallel}}\frac{(q-j_{\parallel})}{2}V_{23} \Bigr] \\ & + z^{2}\Bigl[ -3c_{1}j_{\parallel}^{2} \Bigr] + x\Bigl[ -2c_{2}j_{\perp}\Bigr] + z\Bigl[ -2c_{2}j_{\parallel} \Bigr] - c_{3}.
    \label{eq:app_Res2}
\end{align}
Since both expressions describe the same scalar resistivity field, they must coincide identically. Equating the coefficients of like powers of the spatial coordinates gives the three linear constraints
\begin{align}
	-j_{\perp}(V_{11} + V_{22} + V_{33}) +j_{\perp}\frac{(q+j_{\parallel})}{2}V_{12} - j_{\perp}\frac{(q+j_{\parallel})}{2}V_{21}  + j_{\parallel}(b+2)V_{31} + j_{\parallel}\frac{(q+j_{\parallel})}{2}V_{32} &= 0, \label{eq:vcons_res_1} \\  
	\frac{j_{\perp}}{2}(b-1)V_{12} - j_{\perp}\frac{(q+j_{\parallel})}{2}V_{22} + \frac{j_{\parallel}}{2}(2b+1)V_{32}-\frac{j_{\parallel}j_{\perp}}{2}V_{11}  + \frac{j_{\parallel}}{2}\frac{(q-j_{\parallel})}{2}V_{31}  -\frac{j_{\perp}^{2}}{2}V_{13} -\frac{j_{\perp}}{2}\frac{(q-j_{\parallel})}{2}V_{33} &=0, \label{eq:vcons_res_2} \\  
	-j_{\perp}(b+2)V_{13} + j_{\parallel}(b+1)(V_{11} + V_{22} + V_{33}) -j_{\perp}\frac{(q+j_{\parallel})}{2}V_{23} &= 0. \label{eq:vcons_res_3}
\end{align}
The above relations coincide with Equations~\eqref{eq:vcons_phi_1}--\eqref{eq:vcons_phi_3}, showing that the consistency of the two resistivity expressions is equivalent to the existence of the function $g(x,z)$.

\subsection{Velocity Jacobians}
\label{app:derivation_velocity}

Together with the incompressibility condition,
\begin{equation}
    V_{11}+V_{22}+V_{33}=0,
    \label{eq:vcons_incompressible}
\end{equation}
the three linear constraints obtained above, Equations~\eqref{eq:vcons_res_1} -- \eqref{eq:vcons_res_3}, provide four independent linear relations among the nine elements of the velocity Jacobian. The velocity and magnetic Jacobians must additionally satisfy the stationary momentum equation, Equation~\eqref{eq:st_momentum}. Since both sides of Equation~\eqref{eq:st_momentum} are skew-symmetric matrices, only three independent quadratic relations remain,
\begin{align}
    -j_{\parallel}\left( b+1 \right) -\left( V_{12} - V_{21} \right)\left( V_{11} + V_{22} \right) + V_{23} V_{31} - V_{13} V_{32} & = 0, \label{eq:vcons_mom_1} \\
    -j_{\perp}\frac{\left( q + j_{\parallel} \right)}{2} - V_{12} V_{23} + V_{21} V_{32} -\left( V_{13} - V_{31} \right)\left( V_{11} + V_{33} \right) & = 0, \label{eq:vcons_mom_2} \\
    j_{\perp} - V_{13} V_{21} + V_{12} V_{31} -\left( V_{23} - V_{32} \right)\left( V_{22} + V_{33} \right) & = 0. \label{eq:vcons_mom_3}
\end{align}
Together with the four linear constraints, Equations~\eqref{eq:vcons_res_1}--\eqref{eq:vcons_incompressible}, these equations constitute seven independent constraints on the nine elements of the velocity Jacobian. Consequently, the elements of the velocity Jacobian can be expressed in terms of the elements of the magnetic Jacobian and two independent parameters, which we choose as $V_{11}$ and $V_{13}$. 

The four linear relations are solved first. Incompressibility gives $V_{33}=-V_{11}-V_{22}$, and Equation~\eqref{eq:vcons_res_3} then yields
\begin{equation}
    V_{23}=-\,\frac{2(b+2)}{q+j_{\parallel}}\,V_{13},
    \label{eq:V23sol}
\end{equation}
independently of $j_{\perp}$. Equations~\eqref{eq:vcons_res_1} and \eqref{eq:vcons_res_2} then determine $V_{31}$ and $V_{32}$, in terms of $V_{11}$, $V_{12}$, $V_{13}$, $V_{21}$ and $V_{22}$ Of the three momentum equations, Equation~\eqref{eq:vcons_mom_1} is linear in $V_{22}$ and can be used to eliminate it. The remaining two equations involve only $V_{12}$ and $V_{21}$, together with the free parameters. Eliminating $V_{21}$ by taking their resultant gives a sixth-degree polynomial in $V_{12}$, with coefficients that are polynomials in $V_{11}$ and $V_{13}$.

\subsubsection{The example configuration}
\label{app:example}

The structure of this reduction is most clearly illustrated for the fiducial magnetic field (Equation~\eqref{exB1}) prescribed by the parameters $b=1/2$, $q=3/10$, and $j_{\parallel}=j_{\perp}=1/10$ on the slice $V_{13}=0$, which is the focus of Section~\ref{sec:results}. Equation~\eqref{eq:V23sol} then gives $V_{23}=0$, and the velocity Jacobian takes the block lower-triangular form
\begin{equation}
    \mathcal{V}=
    \begin{pmatrix}
        V_{11} & V_{12} & 0 \\
        V_{21} & V_{22} & 0 \\
        V_{31} & V_{32} & V_{33}
    \end{pmatrix}.
\end{equation}
Equation~\eqref{eq:vcons_mom_1} first yields
\begin{equation}
    V_{22}=-V_{11}-\frac{3}{20\,d},
\end{equation}
where $d=V_{12}-V_{21}$. Substituting this expression into
Equations~\eqref{eq:vcons_res_1} and \eqref{eq:vcons_res_2} then gives
\begin{subequations}
\begin{align}
    V_{31} &=
    \frac{d\left(60V_{11}-500V_{12}+400V_{21}\right)+9}
    {4980\,d}, \\
    V_{32} &=
    \frac{d\left(-300V_{11}+508V_{12}-8V_{21}\right)-45}
    {1992\,d}.
\end{align}
\end{subequations}
Multiplying Equations~\eqref{eq:vcons_mom_2} and
\eqref{eq:vcons_mom_3} by $d^2$ and $d$, respectively, removes the pole and leaves two polynomial equations in $V_{12}$ and $V_{21}$. Eliminating $V_{21}$ gives a degree-six polynomial in $V_{12}$,
\begin{align}
    a_{6}(V_{11})\,V_{12}^{6} + a_{5}(V_{11})\,V_{12}^{5} + a_{4}(V_{11})\,V_{12}^{4} + a_{3}(V_{11})\,V_{12}^{3} +a_{2}(V_{11})\,V_{12}^{2} + a_{1}(V_{11})\,V_{12} + a_{0}(V_{11}) = 0,
    \label{eq:sextic}
\end{align}
where $a_{i}$ are functions of the free parameter $V_{11}$ and can be written as
\begin{subequations}
    \begin{align}
        a_{6} & = 1, \\
        a_{5} & = \frac{101\,131\,V_{11}}{7\,300}, \\
        a_{4} & = \frac{-25\,649 + 1\,006\,878\,V_{11}^{2}}{73\,000}, \\
        a_{3} & = - \frac{7\,V_{11} (1\,286\,061 + 3\,954\,925\,V_{11}^{2})}{730\,000}, \\
        a_{2} & = \frac{-6\,431\,637 + 28\,931\,049\,V_{11}^{2} + 104\,701\,100\,V_{11}^{4}}{7\,300\,000}, \\
        a_{1} & = -\frac{V_{11} (27\,060 + 2\,590\,741\,V_{11}^{2} + 5\,476\,500\,V_{11}^4)}{2\,920\,000}, \\
        a_{0} & = \frac{3 (-165\,336 - 94\,053 \,V_{11}^2 + 986\,200 \,V_{11}^4 + 750\,000 \,V_{11}^6)}{29\,200\,000}.
    \end{align}
\end{subequations}

Two properties of Equation~\eqref{eq:sextic} are important for the numerical analysis. The coefficient of $V_{12}^{6}$ is a nonzero constant, so the polynomial remains sixth order for all $V_{11}$. The polynomial does not factor over the rationals, and no solution in radicals was found. The roots are therefore determined numerically on a $V_{11}$ grid, as described in Sect.~\ref{sec:results}. The resulting dependence of the remaining velocity Jacobian components on $V_{11}$, with $V_{23}=0$, together with $\det\mathcal{V}$ for the four real branches, is shown in Figure~\ref{fig:app_solutions}.

\renewcommand{\arraystretch}{0.6}
\newcommand{\panelw}{0.31\textwidth}
\newcommand{\rowlab}[1]{\rotatebox[origin=c]{90}{\small $#1$}}

\begin{figure*}[p]
  \centering
  \setlength{\tabcolsep}{1pt}
    \begin{tabular}{c@{\hspace{2pt}}c@{\hspace{12pt}}c@{\hspace{12pt}}c}
     & \small $k=1$ & \small $k=2$ & \small $k=4$ \\[2pt]
    \rowlab{V_{12}} & 
      \includegraphics[width=\panelw]{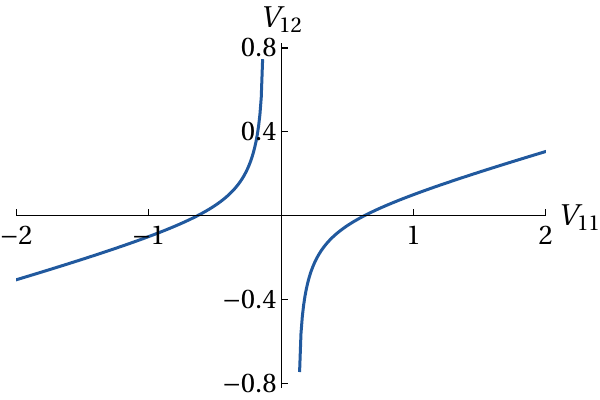} &
      \includegraphics[width=\panelw]{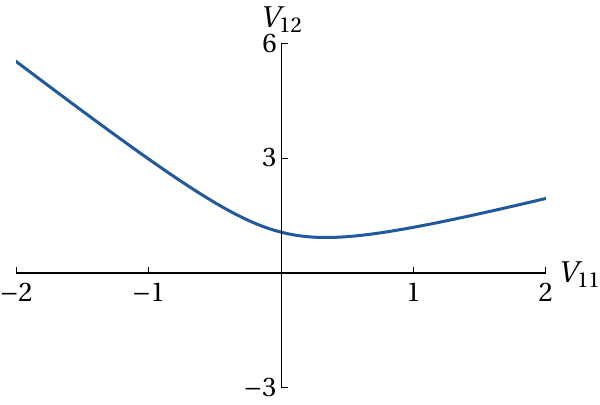} &
      \includegraphics[width=\panelw]{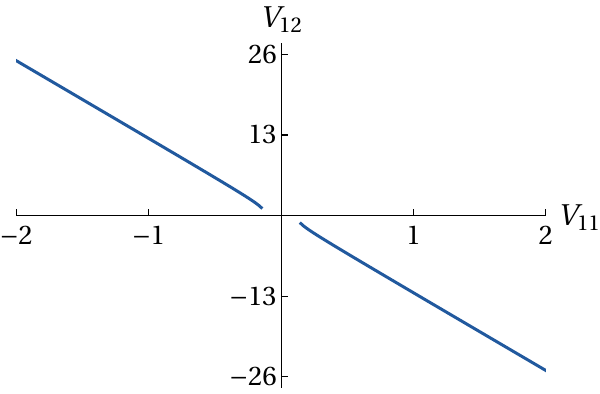} \\[10pt]
    \rowlab{V_{21}} &
      \includegraphics[width=\panelw]{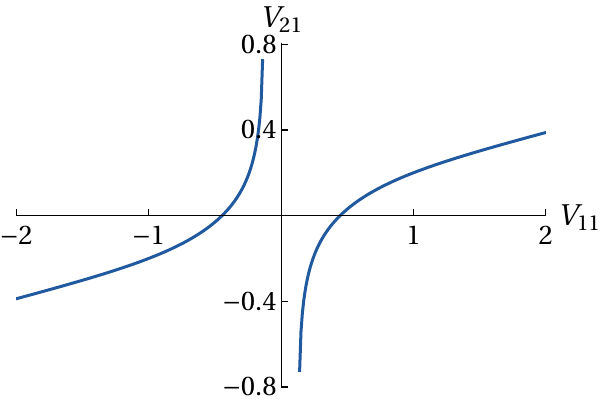} &
      \includegraphics[width=\panelw]{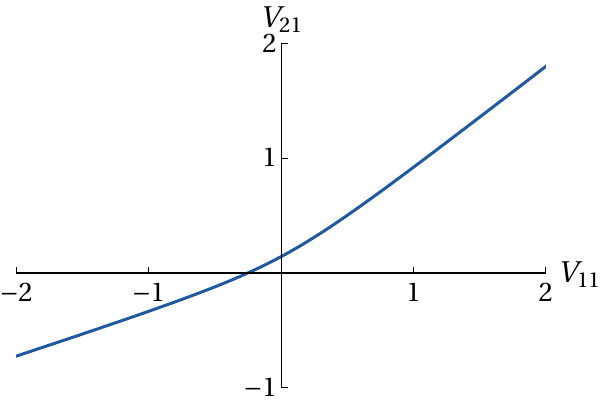} &
      \includegraphics[width=\panelw]{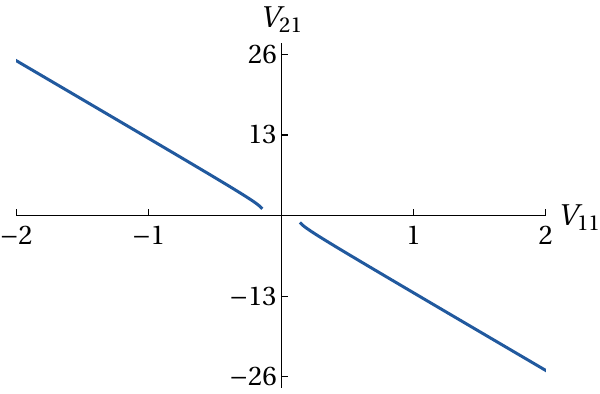} \\[10pt]
    \rowlab{V_{22}} &
      \includegraphics[width=\panelw]{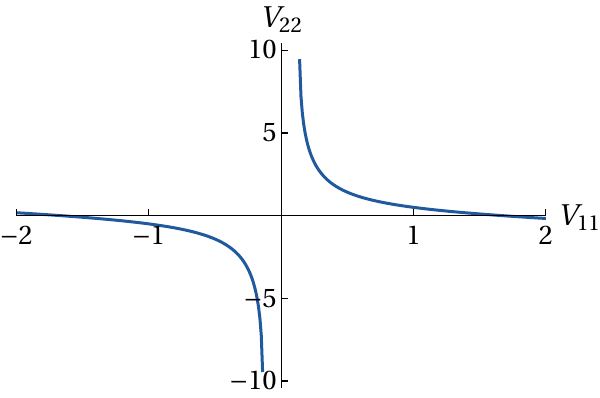} &
      \includegraphics[width=\panelw]{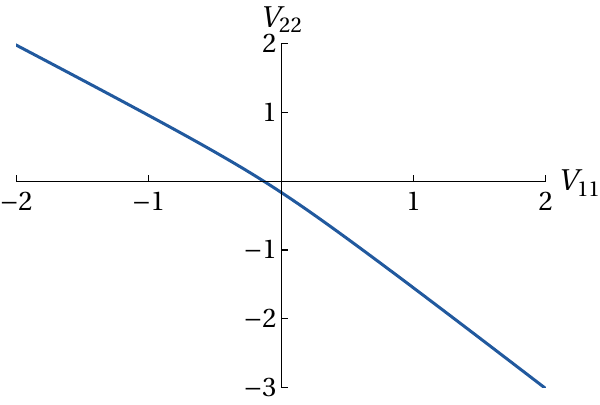} &
      \includegraphics[width=\panelw]{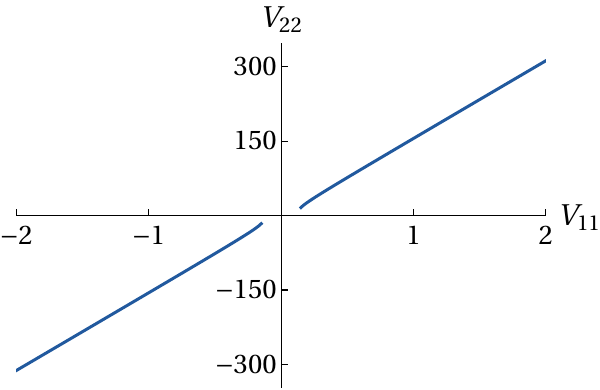} \\[10pt]
    \rowlab{V_{31}} &
      \includegraphics[width=\panelw]{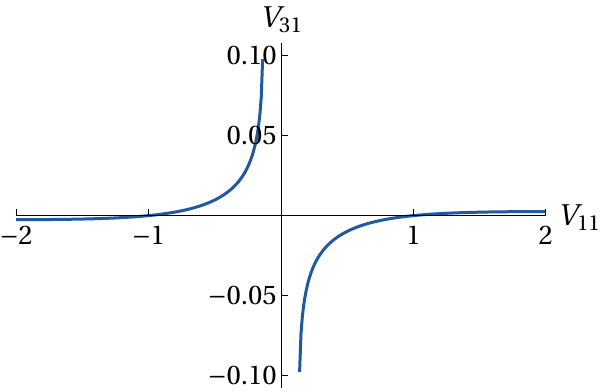} &
      \includegraphics[width=\panelw]{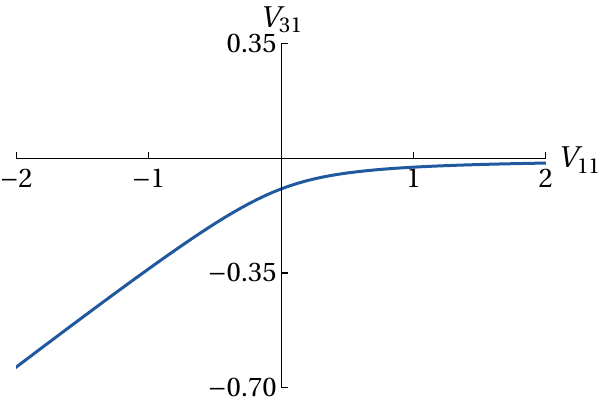} &
      \includegraphics[width=\panelw]{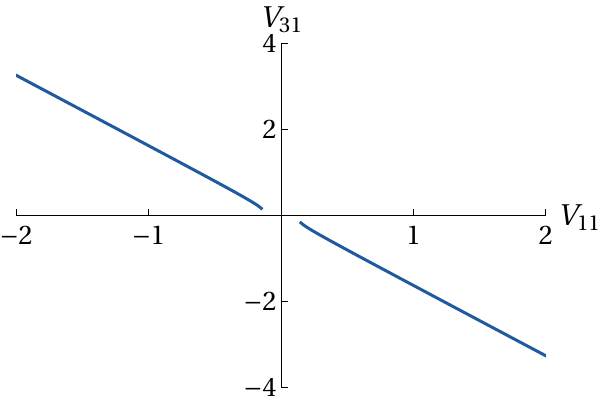} \\[10pt]
      \rowlab{V_{32}} &
      \includegraphics[width=\panelw]{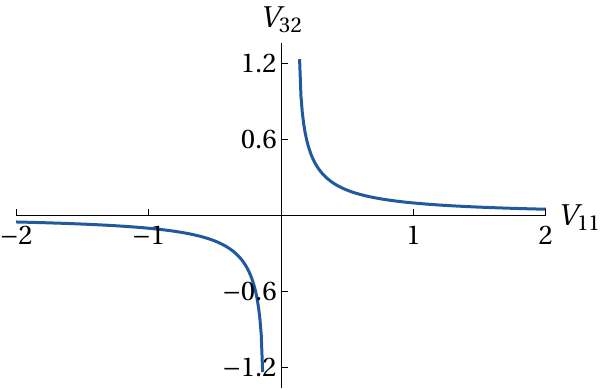} &
      \includegraphics[width=\panelw]{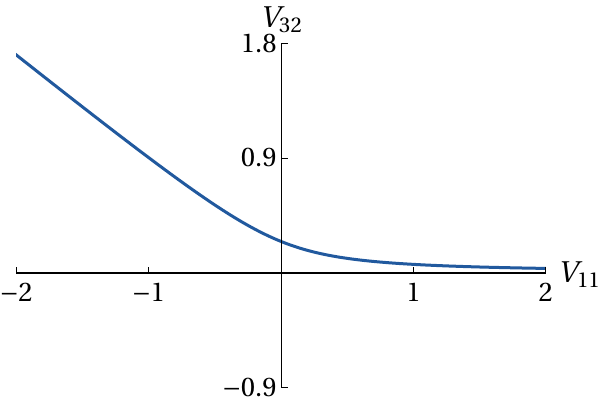} &
      \includegraphics[width=\panelw]{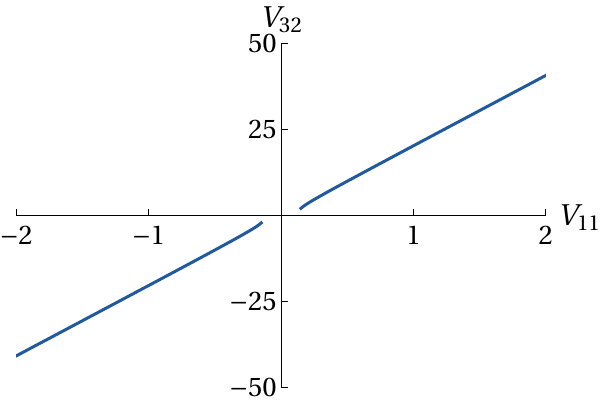} 
  \end{tabular}

  \caption{Elements of the velocity Jacobian and its determinant along the slice
  $V_{13}=0$, as functions of $V_{11}$. The rows correspond to the individual
  components, while the three columns show the solution branches $k = 1,\,2$ and $4$. Gaps indicate intervals in which the branches $k = 1$ and $k=4$ admit no real solution. Note that the vertical scales differ between panels. As the $k=3$ branch is related to the $k=2$ branch by Equation~\eqref{eq:branchsym}, its plots can be obtained by a $180^\circ$ rotation of the $k=2$ plots about the point $(0,0)$ and are therefore omitted here.
  }
  \label{fig:app_solutions}
\end{figure*}

\begin{figure*}
    \ContinuedFloat
    \centering
    \setlength{\tabcolsep}{1pt}
     \begin{tabular}{c@{\hspace{2pt}}c@{\hspace{8pt}}c@{\hspace{8pt}}c}
     & \small $k=1$ & \small $k=2$ & \small $k=4$ \\[2pt]
    \rowlab{V_{33}} &
      \includegraphics[width=\panelw]{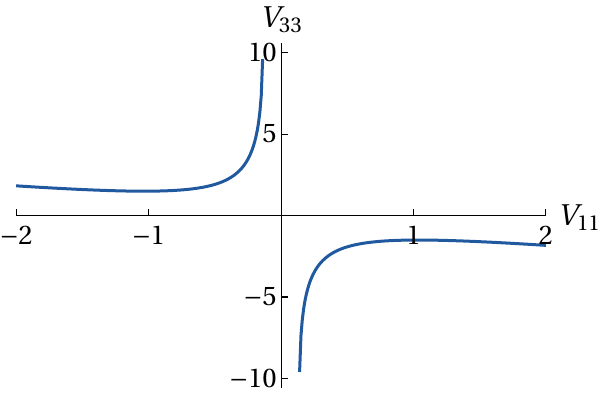} &
      \includegraphics[width=\panelw]{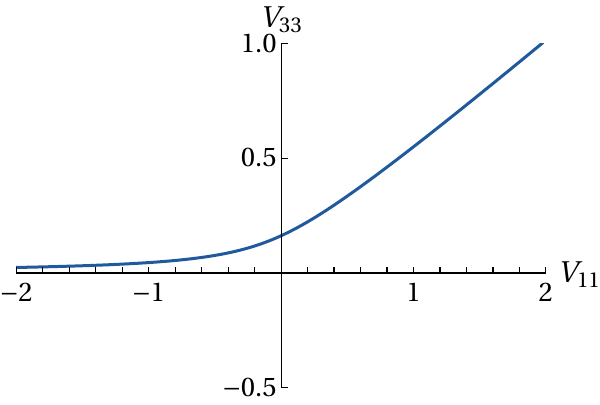} &
      \includegraphics[width=\panelw]{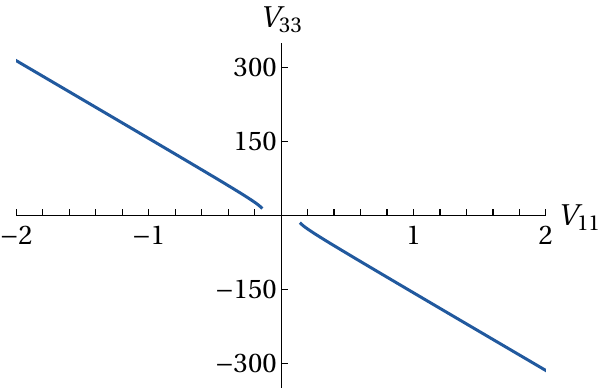} \\[10pt]
    \rowlab{\det\mathcal{V}} &
      \includegraphics[width=\panelw]{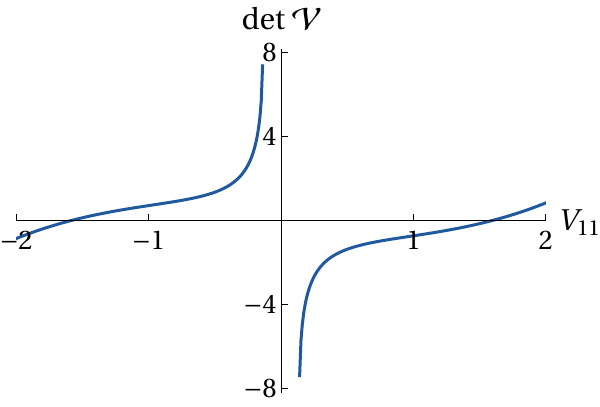} &
      \includegraphics[width=\panelw]{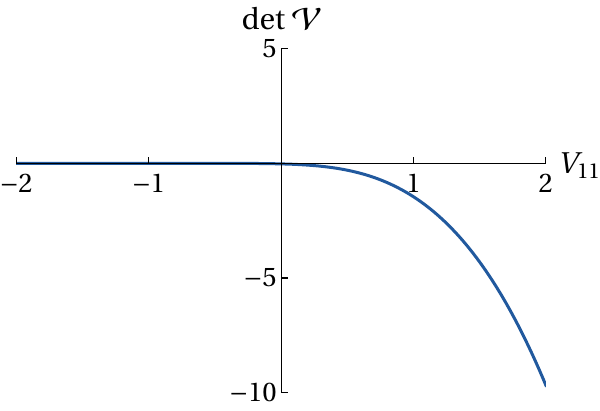} &
      \includegraphics[width=\panelw]{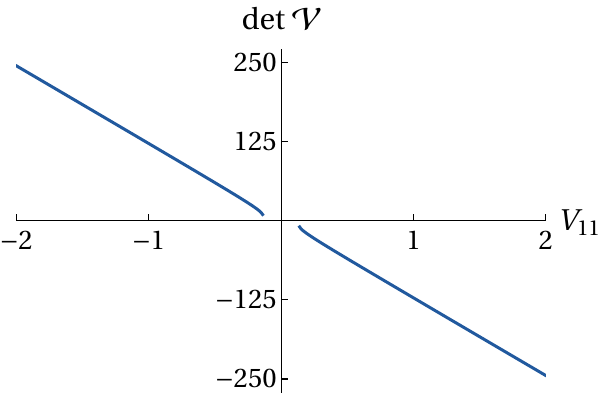} 
     \end{tabular}

    \caption{Continued.} 
\end{figure*}

\end{appendix}

\end{document}